\documentclass[preprint]{aastex}

\usepackage{epsfig,amsmath}

\shorttitle{}
\shortauthors{}

\begin{document}

\title{The Carnegie Hubble Program: The Leavitt Law at $3.6 \mu$m and $4.5 \mu$m in the Large Magellanic Cloud}

\author{\bf Victoria Scowcroft, Wendy L. Freedman, Barry F. Madore, Andrew J. Monson, S. E.  Persson, Mark Seibert,}
\affil{Observatories of the Carnegie Institution of Washington \\ 813
Santa Barbara St., Pasadena, CA~91101}
\author{\bf Jane R. Rigby,}
\affil{NASA Goddard Space Flight Center, Code 665, Greenbelt MD 20771}
\author{\bf \&}
\author{\bf Laura Sturch}
\affil{Astronomy Department, Boston University \\725 Commonwealth Avenue, Boston, MA~02215}
\email{vs@obs.carnegiescience.edu, wendy@obs.carnegiescience.edu, 
barry@obs.carnegiescience.edu, amonson@obs.carnegiescience.edu,
persson@obs.carnegiescience.edu, 
mseibert@obs.carnegiescience.edu,
jane.r.rigby@nasa.gov,
lsturch@bu.edu}

\begin{abstract}
The Carnegie Hubble Program (CHP) is designed to improve the extragalactic distance scale using data from the post-cryogenic era of Spitzer. The ultimate goal is a determination of the Hubble constant to an accuracy of 2\%. This paper is the first in a series on the Cepheid population of the Large Magellanic Cloud, and focusses on the period--luminosity relations (Leavitt laws) that will be used, in conjunction with observations of Milky Way Cepheids, to set the slope and zero--point of the Cepheid distance scale in the mid--infrared. To this end, we have obtained uniformly--sampled light curves for 85 LMC Cepheids, having periods between 6 and 140 days. Period--luminosity and period--color relations are presented in the 3.6~$\mu$m and 4.5~$\mu$m bands. We demonstrate that the 3.6~$\mu$m band is a superb distance indicator. The cyclical variation of the [3.6]--[4.5] color has been measured for the first time. We attribute the amplitude and phase of the color curves to the dissociation and recombination of CO molecules in the Cepheid's atmosphere. The CO affects only the 4.5~$\mu$m flux making it a potential metallicity indicator.
\end{abstract}


\section{Introduction}
\label{sec:intro}
In the past decade, the ``factor-of-two controversy'' surrounding the value of the Hubble constant has been resolved \citep{2001ApJ...553...47F, 2007ApJS..170..377S, 2010ARA&A..48..673F, 2011ApJ...730..119R}. However, the uncertainty on H$_{0}$ currently remains at the level of a few percent and dominates the error budget when deriving other cosmological parameters such as the equation of state parameter, $\omega_{0}$. Reducing the uncertainty on H$_{0}$ will enable us to further constrain the parameters in the concordance cosmological model --- as \citet{2005ASPC..339..215H} states: ``The single most important complement to the CMB for measuring the dark energy equation of state at $z \sim 0.5$ is a determination of the Hubble constant to better than a few percent.''

The Carnegie Hubble Program (CHP) is a ``Warm Spitzer" legacy mission (Spitzer Exploration Program PID 60010 ``The Hubble Constant", P.I. Freedman). It has the primary goal of reducing the systematic uncertainties on the Hubble constant to 3\% using Spitzer data alone, and ultimately 2\% with the addition of JWST. The program  uses the $3.6$ and 4.5~$\mu$m bands of IRAC \citep{2004ApJS..154...10F} to recalibrate the Cepheid distance scale in the mid-infrared, moving out in to the Hubble flow by calibrating and applying a revised mid--IR Tully--Fisher relation. 

A detailed description of the Carnegie Hubble Program is given in \citet{F11_sub}; an overview of our strategy follows below. 

The zero--point of the Cepheid period--luminosity (PL) relation (Leavitt law) will be calibrated using Galactic Cepheids. This sample contains the 10 Cepheids with trigonometric parallaxes measured using the Fine Guidance Sensors on the \textit{Hubble Space Telescope} \citep{2007AJ....133.1810B}. We also observed 27 Cepheids within 4 kpc of the Sun --- close enough for GAIA parallax measurements. Seventeen of these are known members of Galactic open clusters for which main sequence fitting distances are available. The Milky Way program is described in Monson et al. (2011, in preparation).

The second rung on the distance ladder (the subject of this work) involves the long--period ($P>10$ days) end of the Leavitt law. We will calibrate the slope and dispersion of this period range using a sample from the Large Magellanic Cloud. The LMC is an important stepping stone for the extragalactic distance scale; it was used as the base for the Key Project distance scale \citep{2001ApJ...553...47F}, but uncertainty in its distance eventually became the dominating systematic. The calibration found here will then be applied to nearby galaxies in the Local Group, forming a Cepheid distance ladder based on a single telescope and instrument. Additionally, by including galaxies such as M33, the effect of metallicity on the PL slope and zero--point will be studied \citep[e.g.,][]{2009MNRAS.396.1287S}. The calibration will then be applied at a step further out, extending to galaxies beyond the Local Group, such as M81 and NGC 4258, the well known maser galaxy. 

Next, we will extend our distance ladder into the Hubble Flow using the Tully--Fisher relation, itself re-calibrated in the mid--infrared. The dispersion of this relation is expected to drop at IRAC wavelengths, simply due to the decreased sensitivity to inclination--dependent extinction corrections, and will be calibrated using galaxies that contain Cepheids. The distances to over 500 target galaxies in our program will be measured. Finally, we will complete our distance ladder by adding 44 supernova--host field galaxies. These steps will allow us to measure accurate and consistent distances from the closest Cepheids to the most distant galaxies.

This work deals with Cepheids in the Large Magellanic Cloud (LMC). The LMC Cepheid population has been studied extensively for over half a century. For example, \citet{1940AnHar..90..253S} presented well--sampled light curves for 40 Cepheids in the LMC, and study of the population continues to this day with large-scale projects such as OGLE III \citep{2008AcA....58..163S}. Its proximity means we can readily observe Cepheids over a range of periods, from those that are similar to many of the Galactic parallax sample (i.e. $P \leq 10$ days), to the long period Cepheids with $P \geq 10$ days, which more generally overlap Cepheids most easily observed in our most distant targets. 

Until relatively recently, the majority of distance measurements have been undertaken at optical wavelengths (e.g. F01). Cepheid studies at optical wavelengths have their shortcomings, the main one being extinction. Although the effect can, to first order, be removed by use of the reddening--free Wesenheit index \citep{1976RGOB..182..153M} this technique still requires prior knowledge of the extinction law, as well as an assumption that it is universal. By moving to the mid--infrared, reddening and extinction are diminished by around a factor of twenty \citep{1985ApJ...288..618R}, making their absolute contribution and their uncertainties negligible. In addition to the drop in extinction effects, \citet{1982ApJ...257L..33M} describe two other advantages of the infrared over the optical. In the infrared the amplitudes of the Cepheids' lightcurves decrease, as does the intrinsic width of the instability strip, because these wavelengths are less sensitive to temperature changes. Indeed \citet{1982ApJ...257L..33M} demonstrated that the width of the near--infrared $H$ band period--luminosity relation from \textit{random} observations is less than the width from time--averaged $B$--band observations. Near--infrared observations of LMC Cepheids were more recently obtained by \citet{2004AJ....128.2239P}. By moving to longer wavelengths in the mid--infrared, in combination with well phased observations, we can decrease the measured width even further. 

The IRAC imager on \textit{Spitzer} is a superb instrument for undertaking a recalibration of the Cepheid distance scale. Like \textit{Hubble}, \textit{Spitzer} has the advantage of operating in a stable environment without weather or seeing variations, and with great flexibility in scheduling. As such, we have been able to obtain precise and deterministically well--sampled light curves for 85 Cepheids. The observations of the LMC Cepheids were one of the first programs undertaken by post-cryogenic ``Warm Spitzer". 

The target selection and observations are described in Section~\ref{sec:observations} and 
the photometry and calibration are discussed in Section~\ref{sec:data_reduction}. 
Light and color curves are presented for each Cepheid in Section~\ref{sec:results}. PL relations, including a discussion on their use in determining the tilt of the LMC, are given in Section~\ref{sec:pl_relations}. The period--color relation at mean light has been measured for the first time at these wavelengths; it is discussed in Section~\ref{sec:co_absorption}. Section~\ref{sec:summary} provides a summary. 

\section{Observations}
\label{sec:observations}
\subsection{Target Selection}
\label{sec:target_selection}

The LMC has a large sample of well--studied Cepheids to choose from, as light curves have been measured for many thousands. The OGLE studies \citep{1992AcA....42..253U, 1997AcA....47..319U, 2008AcA....58..163S}, while having a primary goal of observing micro-lensing events, have produced extremely large Cepheid catalogs for the LMC and SMC in the optical. For this study, we have chosen a subset of the sample observed by \citet[hereafter P04]{2004AJ....128.2239P}, who studied 92 LMC Cepheids in the infrared $JHK_s$ bands. These Cepheids also had optical ($UBVRI$) light curves available. The majority of them are free from crowding, making them ideal for this study. We chose the 85 Cepheids with periods greater than about 6 days. This ensures that the period distribution overlaps with both the Galactic sample, as well as the most remote galaxies in our program. The Cepheids are distributed over the face of the LMC, allowing any known inclination effects (c.f., P04) to be re-measured and accounted for. 

\subsection{``Warm Spitzer" Observations}
\label{sec:warm_spitzer}
The data were obtained in programs 61000, 61004, 61005, 61006, and 61007, executed between 2009 October 3 and 2010 July 18, at the start of the ``Warm Spitzer" mission. Each Cepheid was observed at 24 epochs. For those Cepheids with $P > 12$~d, the observations were taken over a single cycle, spaced on average by P/24 days.  This produced regularly--sampled light curves while preserving the high scheduling efficiency of the spacecraft. The shortest--period Cepheids (P $\le$ 12~d) could not be observed in such a tightly--phased manner; instead we scheduled  24 observations separated by  $12 \pm 4$~days. The Appendix explains the choice of 24 evenly-spaced epochs. 

Each IRAC observation consisted of ten frames, five in each of the 3.6~$\mu$m and 4.5~$\mu$m bands, with a medium-scale, 5-point gaussian dither pattern. (Dithering of exposures is essential for accurate photometry.) The frame time was 2.0~s, giving an effective per-pixel integration time of 1.2~s in each frame. IRAC is known to be susceptible to persistent images (i.e. latent images of bright, but often unsaturated, stars can remain in later frames) and this effect is apparently worse in Warm Spitzer than in the original mission (see IRAC handbook, version 2.0\footnote{\url{http://irsa.ipac.caltech.edu/data/SPITZER/docs/irac/iracinstrumenthandbook/}}, section 7.2.8). By using a large scale dither pattern, where subsequent observations of the Cepheid are spaced by approximately half the chip, we have mitigated the possibility of our photometry being affected by previous observations. 

\section{Data reduction and Photometry}
\label{sec:data_reduction}
These early ``Warm Spitzer" data were some of the first to be reduced with the new ``Warm IRAC" pipeline. The images were processed using the Spitzer Science Center pipeline version S18.18.0, which does not include the artifact correction steps present in the cold mission pipelines. Therefore, in addition to the normal Spitzer Science Center (SSC) processing, the BCD (Basic Calibrated Data) frames were run through the Warm--Mission Column Pulldown Corrector\footnote{\url{http://ssc.spitzer.caltech.edu/dataanalysistools/tools/contributed/irac/fixpulldown/}}. {\sc mopex} \citep{2005ASPC..347...81M} was used with the standard {\it overlap} and {\it mosaic} name lists for the overlap correction and mosaicking.

The most important objective of this work is to produce an accurate calibration of the Cepheid period--luminosity relation. Our photometry is on the same scale as the standard calibration described in \citet{2005PASP..117..978R}. They define the flux scale and zero point such that they correspond to a 10-pixel radius aperture, with a sky annulus running from 12 to 20 pixels (in units of native IRAC pixels) with pixel--phase and array--location corrections applied as necessary. 

Although the majority of the Cepheids in our sample are isolated, several have close neighbors that fall inside the 10-pixel aperture. These stars could not be measured correctly using aperture photometry. In order to reduce the entire LMC sample in a consistent way, Point Response Function (PRF) fitting  was used in place of aperture photometry, utilizing the SSC {\sc apex} software \citep{2005PASP..117.1113M}. The point response function (PRF) models the detector response to a point source, and can be thought of as an over-sampled representation of the point spread function. The BCDs were mosaicked to obtain the bad pixel maps, but the PRF fitting was performed on each BCD individually using the {\it apex\_1frame} name list, with two changes. First, rather than using the PRF corresponding to the center of the array, the set of 25 spatially--dependent PRFs was used for each channel\footnote{PRFs are available from \url{http://ssc.spitzer.caltech.edu/irac/calibrationfiles/psfprf/}}. The IRAC detectors are known to have some spatial response variations; using the PRF map ensured that any variation in flux was properly accounted for. Second, the normalization radius -- the radius to which the fit is calculated -- was set to 1000 pixels. As the PRF is sampled at 100 times per pixel, the 1000 pixel normalization radius corresponds to the calibration aperture radius. For each BCD, all the stars in each image were measured. 

\subsection{PRF calibration}
\label{sec:prf_calibration}
The construction of the IRAC PRF is a complex task requiring a large number of observations of a stable star. Warm Spitzer PRF models are not yet available from the Spitzer Science Center; however, it is still possible to obtain accurate photometry using the cold PRF, correcting as necessary for the predicted differences between the cold and warm models. The most significant difference between the two models is the pixel--phase correction (PPC).

The PPC is necessary given that the response of the IRAC detectors is known to vary on a sub--pixel level. This means that the measured flux of an object not only depends on the position of the star in the frame, but also on the position of the peak of the PSF within the pixel. The fractional position within a pixel is known as the ``pixel phase". In the cryogenic mission the pixel--phase effect was around 4\% (peak--to--peak), but is approximately double that in the warm mission. In addition to the change in magnitude, whereas the cold mission PPC was represented by a symmetric Gaussian centered on the peak response position, the warm mission PPC is described by a Gaussian, differing in width in the $x$ and $y$ directions.

As the PRF is an oversampled representation of the pixel's response to light, it includes the cold mission PPC by definition. To correct the warm mission data, the cold PPC must first be multiplied out of the {\sc apex} photometry results, and the warm correction then divided in. Although it is possible that the pixel--phase effect will cancel out when the results from a dither pattern are averaged together, in our data the PPC was applied to each BCD before the averaging took place. 

\subsection{Photometric stability}
\label{sec:stability}
\begin{figure}
\begin{center}
\includegraphics[width=150mm]{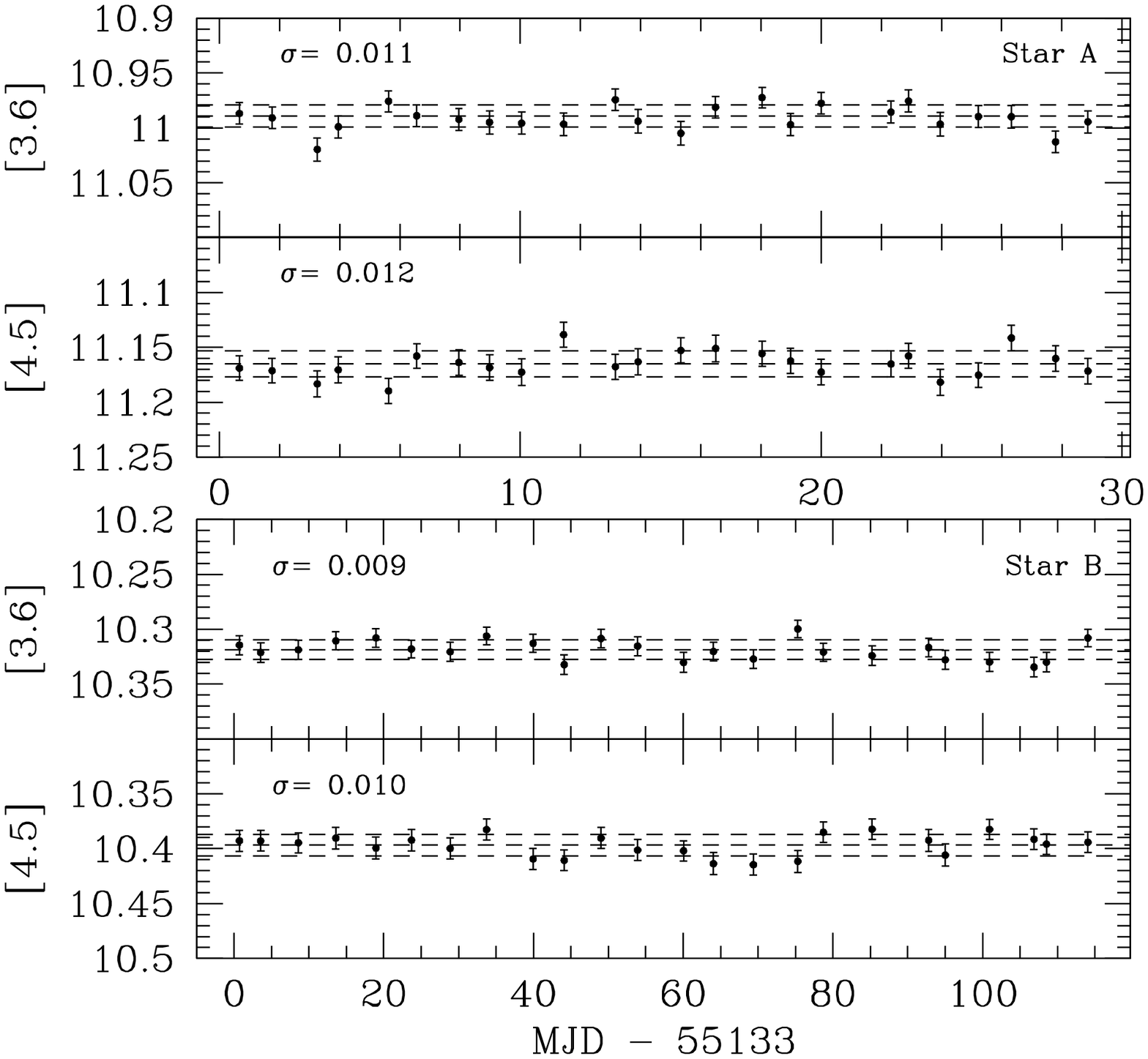}
\caption{Photometry of two randomly chosen constant stars in the HV00872 (Star A) and HV02883 (Star B) fields. The dashed lines show the average magnitude of the star and $\pm1\sigma$ limits. The dispersion in both cases is approximately 0.01 mag. The time baselines differ as the Cepheids in the fields were observed with different cadences. No systematic trends are seen.}
\label{fig:constant_stability}
 \end{center}
\end{figure}

The stability of IRAC is of fundamental importance to our program. We have quantified the stability of the instrumental zero--point by measuring constant stars in our Cepheid fields, and monitoring their magnitudes over time. The photometry method was identical to that used on the Cepheid targets. In Figure~\ref{fig:constant_stability} we show the photometry for constant stars in two fields. Star A was chosen in the HV00872 field, at $\alpha = 4^{h}55^{m}17.09^{s}$, $\delta = -67^{\circ}27'53.82"$, and Star B was in the HV02883 field at $\alpha = 4^{h}56^{m}21.24^{s}$, $\delta = -64^{\circ}41'35.65"$.  In both cases, the dispersion of the measurements is approximately $\pm0.01$ mag, and no systematic trend with time is seen. The Cepheids in the fields have very different periods (29.82~d vs. 108.97~d), hence the time baseline and the cadence of the observations differ. This confirms both the short and long term stability of IRAC.

\subsection{Photometry comparison}
\label{sec:prf_phot_testing}
As mentioned above, the IRAC magnitude system is defined by the flux measured within a 10-pixel aperture. To verify that the photometry for this work is on the calibrated system, the data for the Cepheid HV00872 were also processed using the SSC IDL routine {\it photo\_for\_star.pro} (SSC 2010, private communication). This routine performs aperture photometry on a single star using two aperture sizes --- a 3 pixel aperture with a 3--7 pixel sky annulus, and the 10 pixel aperture with a 12--20 pixel sky annulus used in the calibration definition given by \citet{2005PASP..117..978R}. It also includes the latest pixel--phase corrections and array--location corrections.

The {\sc apex} photometry was compared to the results from both aperture sizes. No systematic trends with magnitude were found. The difference between the {\sc apex} and the 3-pixel aperture fluxes were consistent with the published aperture correction for transformations between the 3 and 10 pixel apertures. Both the {\sc apex} and small aperture results were then compared with the large aperture values. As was expected, the scatter in the large aperture results was much larger than the other two methods, but no systematics were found. Aperture corrections were derived for each band, and agreed with the cold mission values (1.124 and 1.127 at 3.6~$\mu$m and 4.5~$\mu$m respectively) to within 1\%. 

To put the PRF photometry on the same system as the standard aperture photometry the following correction was used:
\begin{equation}
F = \dfrac{F_{PRF} \times A \times PPC_{Cold}}{PPC_{Warm}}
\label{eqn:flux_corrector}
\end{equation}
where $F$ is the flux in the standard system, $F_{PRF}$ is the flux output by {\sc apex}, $A$ is the aperture correction (1.03 and 1.02 for channels 1 and 2 respectively) and $PPC_{Cold}$ and $PPC_{Warm}$ are the cold and warm pixel phase corrections. The fluxes were converted to magnitudes using the standard zero--magnitude flux densities of 280.9~Jy~([3.6]) and 179.7~Jy~([4.5]).

\section {Results}
\label{sec:results}
\subsection{Periods}
\label{sec:periods}
We first refined the available periods as follows. Light curves were produced in the 3.6 and 4.5~$\mu$m bands. The data points from \citet{2009ApJ...695..988M}, taken as part of the SAGE project \citep{2006AJ....132.2268M}, were also phased in. The periods given in P04 were taken as initial values; these were improved by iterating on the period and re-phasing the data until a minimum dispersion in the light curves was obtained in all available wavebands ($U,B,V,R,I,J,H,K,[3.6],[4.5],[5.8],[8.0]$, but note that not all wavelengths were available for all Cepheids). On average, the periods changed by less than 1\%. In some cases it was clear that the period of the Cepheid had changed in the time interval between the optical and infrared observations. This is to be expected, as we utilized as much archival data as possible; the time baseline for some Cepheids was of the order of many decades, which is comparable to the time many Cepheids have been found to significantly change their periods \citep[for a discussion of period changes in a sample of LMC Cepheids, see][]{2001AcA....51..247P}. The practice of fitting all wavebands simultaneously makes it easy to find objects for which this is the case, as the period that fits the later observations will result in an incorrectly phased light curve for the archival data, and vice versa. When this was the case the IRAC and $JHK$ data alone were used to refine the period and to phase the data. The finally adopted periods are given in Table~\ref{tab:mag_results}. 

\subsection{Infrared Light Curves and Mean Magnitudes}
\label{sec:light_curves}

Once the periods had been adjusted, the IRAC light curves were plotted using just the CHP data. This avoids issues such as slight period changes and larger photometric uncertainties in the archival data that could affect the IRAC fit. The resulting light curves are shown in Figure~\ref{fig:light_curves} in order of increasing period. The {\sc gloess} program (described in section 3.2 of P04) was used to interpolate between the data points and to thus draw the smooth curves. {\sc gloess} was also used to determine time--averaged (intensity--mean) magnitudes and uncertainties. The results are given in Table~\ref{tab:mag_results}.

For the $[3.6]$ (top) and $[4.5]$ (middle) panels, the statistical uncertainties on the individual measurements are comparable to the size of the plotted points. Table~\ref{tab:example_photometry_hv872} gives a sample of the individual data points for HV00872; all data for this program are available with the online version of the article.

The bottom panels of the plots in Figure~\ref{fig:light_curves} show the $[3.6]-[4.5]$ color variation through the Cepheid pulsation cycle. As light curves of this quality have not previously been available in the near-IR, this cyclical effect has not been noted before. The effect is almost certainly due to absorption by the 4.65~$\mu$m CO bandhead in the 4.5~$\mu$m filter, as first suggested by \citet{2010ApJ...709..120M}. The color variations are discussed below in Section~\ref{sec:color_curves}.

\section{Mid-infrared Period--Luminosity Relations}
\label{sec:pl_relations}
Period--luminosity (PL) relations for the two IRAC bands were fitted using the weighted--least--squares technique described in \citet{1996ApJ...470..706A}. This takes into account the intrinsic width of the PL relation, giving a more robust result than a regular weighted--least--squares fit.

The PL relations take the form 
\begin{equation}
M = a (\log P - 1) + b
\label{eqn:model_pl}
\end{equation}
and were fit using four different period cuts. The results of the different fits are given in Table~\ref{tab:pl_relations}. The relations derived from the two phase points available from SAGE \citep{2009ApJ...695..988M} are also given for comparison. In Table~\ref{tab:pl_relations}, $N$ is the number of Cepheids in the fit, $a$ and $b$ were both found using the \citet{1996ApJ...470..706A} weighted--least--squares method, $\sigma_{a}$ and $\sigma_{b}$ are their respective uncertainties, and $\sigma$ is the standard deviation of the fit. 

The finally adopted relations are plotted in Figure~\ref{fig:pl_relations}. The solid line shows the adopted PL relations; these were the fits that used only Cepheids in the range $10$ d $\leq P \leq 60$ d. Note the systematic deviation of the longest--period Cepheids, which are well known to be significantly fainter than the PL relation at virtually all wavelengths. Consistent with other studies (such as F01) we have excluded objects with $P>60$ days from the fitting. The SAGE PLs (given in Table~\ref{tab:pl_relations}) are entirely consistent with the values found in this work. Although we find no evidence in our data to suggest a break or change in slope in the $P=10$ days region (c.f., \citet{2004A&A...424...43S}), we have also cut the low end of the period distribution, again consistent with F01. 

In both cases, the dispersion around the relation has fallen dramatically in comparison with previously published determinations. The previous measurements from SAGE gave standard deviations of 0.135 and 0.141 mag in [3.6] and [4.5] respectively, whereas the scatter around our final relations are 0.108 and 0.115 mag. This is because the magnitudes used in the \citet{2009ApJ...695..988M} PL relation are not fully phase--averaged, but rather they are the simple mean of two random phase points (see the Appendix). 

The precision of our phase--averaged light curves is best demonstrated by comparing the residuals from the [3.6] PL relation to those from the [4.5] relation. This is shown in Figure~\ref{fig:pl_residuals}. In this figure the small inset plot shows the entire data set, with the main plot showing just the objects with $\Delta$ mag $\leq \pm 0.2$, which encompassed 80 of the 85 Cepheids in the sample. It is clear that the residuals are highly correlated. That is, if a star lies above the mean PL relation in one band, it will tend to be above the mean in the other band as well. The same effect in the $J$ and $K_{S}$ bands was found by P04. 

Excluding two outliers, the range in $\Delta$ mag is $\pm 0.2$ mag. This spread is a combination of two effects: depth and tilt effects in the structure of the LMC, and the intrinsic magnitude width of the instability strip. The residual dispersion in this correlation is only 0.02 mag, suggesting that we are seeing the photometric limit of the data.

Figure~\ref{fig:colour_mag_residuals} shows how the residuals of the $([3.6]-[4.5])$ color correlate with the PL residuals of the $[3.6]$ magnitude. As we know from the previous plot that any dispersion in Figure~\ref{fig:colour_mag_residuals} arises from the intrinsic properties of the Cepheids rather than photometric uncertainties, the width of the instability strip can be measured from this figure. Assuming that the instability strip is represented by a rectangular distribution, i.e. it is uniformly filled and has hard limits at the blue and red edges, the dispersion around the mean is related to the width by the relation $\sigma = R / \sqrt{12}$, where $R$ is the full width of the distribution. The color dispersion of the data is $\pm0.018$ mag, leading to a color width of the instability strip of 0.062 mag. 

\subsection{Tilt of the LMC}
\label{sec:tilt}

By combining our knowledge of the Cepheid's location within the LMC with its position in the period--luminosity plane, the tilt of the galaxy with respect to the plane of the sky can be constructed, as was done in P04. 

In Figure~\ref{fig:tilt_plot} we show the positions of the Cepheids within the LMC in projection on the sky. The bottom left panel shows the position of each of the Cepheids relative to the center of the LMC. Solving for PL coefficients $a$ and $b$, but now including $\xi$ and $\eta$ (distances from the center of the LMC in RA and Dec) as free parameters, we can obtain the tilt and the angle of the line of nodes if the Cepheids on average lie in a plane. The top left and bottom right plots show the distribution of the points when viewed from orthogonal directions in the plane of the sky. Empty circles represent the Cepheids that are brighter than the average (i.e. they lie above the mean PL relation), and filled circles show those that are fainter than the average (below the mean PL relation). In this figure the reference coordinate system has been rotated by 50$^{\circ}$  so that the line of nodes of the LMC is parallel with the $x$--axis. In this plot, it is clear that the Cepheids in the bottom half tend to be fainter than the PL, hence further away, while those in the top half tend to be systematically brighter, hence closer than average. The simplest interpretation is that the LMC is in fact tilted with respect to the plane of the sky.

By tilting the LMC by $28^{\circ}$ relative to the plane of the sky we can minimize the width of the histogram in the top right corner of Fig.~\ref{fig:tilt_plot}. This is in good agreement with the results of P04, and will be the subject of a more complete discussion of the data at a later time. 

\section{The Role of CO Absorption}
\label{sec:co_absorption}

We believe that the variations in the $[3.6]-[4.5]$ color, both at mean light and throughout the pulsation cycle, are due to CO absorption in the 4.5~$\mu$m band. We do not see any evidence of extended emission surrounding the Cepheids in our images. This is consistent with the results of \citet{2010ApJ...709..120M, 2010ApJ...725.2392M}  and \citet{2011AJ....141...42B} who found evidence for circumstellar emission in the longer wavelength IRAC bands, but nothing at 3.6 or 4.5~$\mu$m.  \citet{2010ApJ...709..120M} notes that the $[3.6]-[4.5]$ color of the Cepheids is indistinguishable from their non--variable control stars, and is therefore caused by an intrinsic feature of stars of this spectral type, regardless of whether or not they are Cepheids. This rules out circumstellar envelopes and mass--loss as the source of the feature, leaving only the CO bandhead.

\subsection{Period--Color Relation}
\label{sec:pc_relation}

The possibility of a period--color (PC) relationship at mid--infrared wavelengths was correctly noted by \citet{2010ApJ...709..120M}. However, their magnitudes (and therefore colors) were measured at only a single epoch, introducing large scatter into the PC relation, as is shown in their figure 5.  Figure~\ref{fig:pc_relation} shows the relationship between period and $[3.6]-[4.5]$ color for our sample. The period--color relation was fit using Cepheids with $1 \leq \log P \leq 1.8$, and is found to be
\begin{equation}
[3.6] - [4.5] = -0.087 (\pm 0.012) (\log P - 1) + 0.005 (\pm 0.005)\text{.}
\label{eqn:pc_relation}
\end{equation} 
The LMC Cepheids clearly exhibit a period--color relation in this period range. The standard deviation around the relation is $\pm 0.018$ mag. The relation appears to invert for the longest---period ($\log P > 2$) Cepheids. Ignoring the $\log P > 2$ sample, the long period Cepheids are brighter and bluer in $[3.6]-[4.5]$ than their short period counterparts. At longer periods the Cepheids are at cooler average temperatures, and so the carbon and oxygen are situated in CO molecules. As discussed previously, the CO absorbs some of the 4.5~$\mu$m flux, making the star appear bluer. As we move to shorter periods the average temperature rises, the CO is gradually dissociated, and the average colors become redder. The period--color relation reverses at $P>100$ days. We do not know what causes this reversal, but the Cepheids at these periods are deviant in their luminosities and colors at almost all wavelengths.

\subsection{$[3.6]-[4.5]$ Color Curves}
\label{sec:color_curves}

The period-color relation is closely related to the variations in color shown in Figure~\ref{fig:light_curves}, as they are both due to CO affecting the 4.5~$\mu$m flux. In a static stellar atmosphere, with a temperature and surface gravity corresponding to an F--type supergiant, the 4.5~$\mu$m flux is suppressed. In the case of Cepheids, however, we are not dealing with static atmospheres. As the Cepheid contracts its temperature increases. As the star becomes hotter, CO in the atmosphere dissociates, causing less absorption at 4.5~$\mu$m, making the star then appear redder than average. Once the Cepheid begins to expand again its temperature drops, allowing the C and O to recombine. As more CO is present, the 4.5~$\mu$m flux is once again suppressed, making the star bluer, in the particular color combination ($[3.6]-[4.5]$) discussed here.  Inspection of Figure~\ref{fig:light_curves} shows that the phasing of the light and color curves exhibits precisely this behavior. 

There also appears to be a relation between period and the amplitude of the color variation, such that the short period Cepheids ($P \leq 10$ days) tend to have flat color curves. The cyclical variation appears to ``turn on" around $P=10$ days, increasing in amplitude with period. Again, this is a temperature effect. The short period Cepheids are so hot that the CO is always dissociated, whereas the longest period Cepheids have lower maximum temperatures and so never dissociate all their CO, thus have the largest color amplitude. 

The dependence of color on CO implies that the 4.5~$\mu$m band should be subject to a metallicity effect, effectively ruling it out as a distance indicator. Fortunately, the bandhead is confined to the [4.5] filter alone; no CO or other molecular feature is present at 3.6~$\mu$m, and so it can still be confidently used to measure Cepheid distances. The physics behind the cyclical CO variation, along with the effect of metallicity on Cepheid's IRAC colors will be described in detail in Scowcroft et al. (2012, in preparation).

\subsection{PL Slopes and CO Absorption}
\label{sec:pl_slopes}

We now return to consideration of the slopes of the PL at different wavelengths. Figure~\ref{fig:pl_slopes} shows  the dependence of the slope on wavelength, a plot that is repeated from figure 4 of \citet{2008ApJ...679...71F}. The $[3.6]$ and $[4.5]$ slope values have been updated using the Table 3 values in this paper. The slope of the $[4.5]$ PL relation is clearly deviant. We can again understand this as due to CO as follows: as we move from shorter to longer periods we are moving to cooler stars; these have more CO absorption on average. More CO means depressed 4.5~$\mu$m flux and thus a shallower PL slope. Therefore, this plot gives further evidence that the color curves are due to CO absorption that affects only $[4.5]$; the $[3.6]$ magnitude is unaffected. 

A smaller magnitude effect is also seen in the $K$ band where a shallower CO bandhead is present. However, this effect can be mitigated using the $K_{S}$ band, which excludes the longest wavelength section of the bandpass --- the part that happens to include the CO bandhead.

\section{Summary}
\label{sec:summary}
We have used the IRAC camera on ``Warm Spitzer" to measure [3.6] and [4.5] light curves for 85 Cepheids spread across the face of the LMC and covering a 6 to 140 day period range.  Each star has been measured at 24 phase points. Our data agree well with the random phase data of the SAGE project, the latter taken for different purposes. 

The light curves (shown in Figure~\ref{fig:light_curves}) allow us to measure the mean magnitudes and color of each Cepheid to a precision never achieved before in the mid-infrared. These lead to our best PL relations
based on the measurements of the [3.6] and [4.5] magnitudes for 67 stars with $10 \leq P \leq 60$ days. The PL relations have been measured to such high precision that their scatter directly correlates with the front-to-back geometry of the sample of Cepheids in the LMC. This correlation recalls a similar effect presented in P04. Using the residuals of the period--luminosity relation, we have measured the tilt of the galaxy as 28$^{\circ}$, with respect to the plane of the sky. 

The dense and uniform sampling of our Cepheid light curves reveals a cyclical color variation which we interpret as being due to the effect of temperature--sensitive molecular absorption of CO in the Cepheid's atmosphere. When the Cepheid is cool, the CO present in its atmosphere absorbs at 4.67~$\mu$m, suppressing flux exclusively in the IRAC 4.5~$\mu$m band, making the star appear blue in the $[3.6]-[4.5]$ color. As the Cepheid contracts, and so heats up, the CO molecules begin to dissociate, ending the suppression of the $4.5\mu$m flux and reddening the color. The temperature dependence of the CO absorption also explains the period--color relation we observe (see Figure~\ref{fig:pc_relation}): the cooler, longer period Cepheids appear bluer (more highly absorbed by CO at 4.5~$\mu$m) than their short period counterparts. The implications of this temperature--sensitive effect will be studied in detail in Scowcroft et al. (2012, in preparation). 

The period--luminosity relations measured here will be used in combination with those from our Galactic sample (Monson et al. 2011, in preparation), to establish the bottom rung of the extragalactic distance ladder. They will then be combined with parallel observations of Cepheids in the Local Group and beyond, to establish the distance scale to a systematic accuracy not achievable before now. We will also be testing for the effect of metallicity on the mid--IR period--color relation, by combining this sample with Galactic and SMC Cepheids, (Scowcroft et al. 2012, in preparation). 

\section{Acknowledgements}
\label{sec:acknowledgements}
This work is based in part on observations made with the Spitzer Space Telescope, which is operated by the Jet Propulsion Laboratory, California Institute of Technology under a contract with NASA. Support for this work was provided by NASA through an award issued by JPL/Caltech. This research has made use of the NASA/IPAC Extragalactic Database (NED) which is operated by the Jet Propulsion Laboratory, California Institute of Technology, under contract with the National Aeronautics and Space Administration. We would like to thank the staff of the Spitzer Science Center, and in particular Nancy Silbermann, for their assistance with scheduling such a large and complex project. Without their help, this project would not have been possible.

\section{Appendix}
\label{sec:appendix}
The observing plan we adopted using \textit{Spitzer} to optimally follow up known Cepheids is very differently constrained from what one would or could attempt from the ground. Ground-based observations are obviously restricted by 24-hour resonances with the day-night observing cycle. They are often additionally modulated by the lunar cycle (especially for optical observations). They are stochastically interrupted by weather, instrument availability, technical problems, and time assignments. 
Randomly spaced observations of known Cepheids may thus be the only way to fully sample light curves and to determine their mean values. In the case of space-based observations of known Cepheids randomly sampled
data may be the only kind available (e.g. data from the SAGE project \citep{2006AJ....132.2268M}, which was mined for Cepheids by \citet{2008ApJ...679...71F} and \citet{2009ApJ...695..988M}). 

However, the 24-hour availability of a solar-orbiting telescope opens up a unique opportunity to design the observing program to be optimally scheduled on a star-by-star basis. Let us first consider the case of random sampling. In randomly sampling any given distribution function the error on the derived mean value goes down in inverse proportion to the square root of the number N of randomly timed observations (i.e., $\sigma \propto 1/\sqrt{N}$).  What is not widely appreciated is that convergence of the error on the mean can be very significantly accelerated by deterministic (non-random) sampling. In the case of an individual Cepheid where the time distribution of luminosities (i.e., its amplitude) is bounded (but not {\it a priori} known), and where the cycling time (i.e., its period, but not necessarily its phase) is also known, then it is possible \citep[see][]{2005ApJ...630.1054M} to construct a simple observing strategy where the errors on the mean decrease in direct inverse proportion with the number of observations (i.e. $\sigma \propto 1/N$). Not surprisingly, for a specified number of observations N, and a known period of variation P, the optimal separation of data points in time is the uniform distribution P/(N+1).

The distribution function in this case is closely approximated by a rectangular probability density distribution. The equivalent sigma for such a distribution is $A/\sqrt{12}$, where $A$ is the total range of the distribution function; in the case of a Cepheid, $A$ is its amplitude.  In the mid-IR the amplitudes of Cepheids can reach 0.8~mag (e.g. HV00883), which converts to an equivalent sigma of $\pm0.23$~mag.  If the required precision on a determination of the mean magnitude were to be set at 0.01~mag (i.e., one percent) then random sampling of the light curve would require over 500 observations! To reach the same precision using deterministic (uniform) sampling a grand total of 23 observations are required.  To compensate for slight variations away from  exact scheduling we requested and obtained 24 observations per LMC Cepheid. The light curves shown in Figure~\ref{fig:light_curves} are a testament to the care and efficiency with which \textit{Spitzer} was scheduled for this program.

\label{references}

\begin{figure} 
 \begin{center}$ 
 \begin{array}{ccc} 
\includegraphics[width=50mm,angle=270]{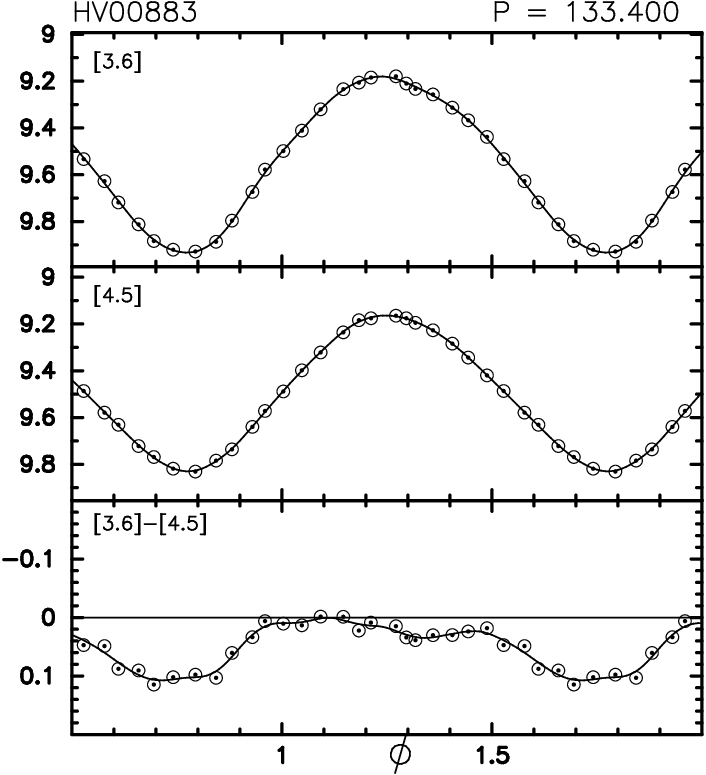} &
\includegraphics[width=50mm,angle=270]{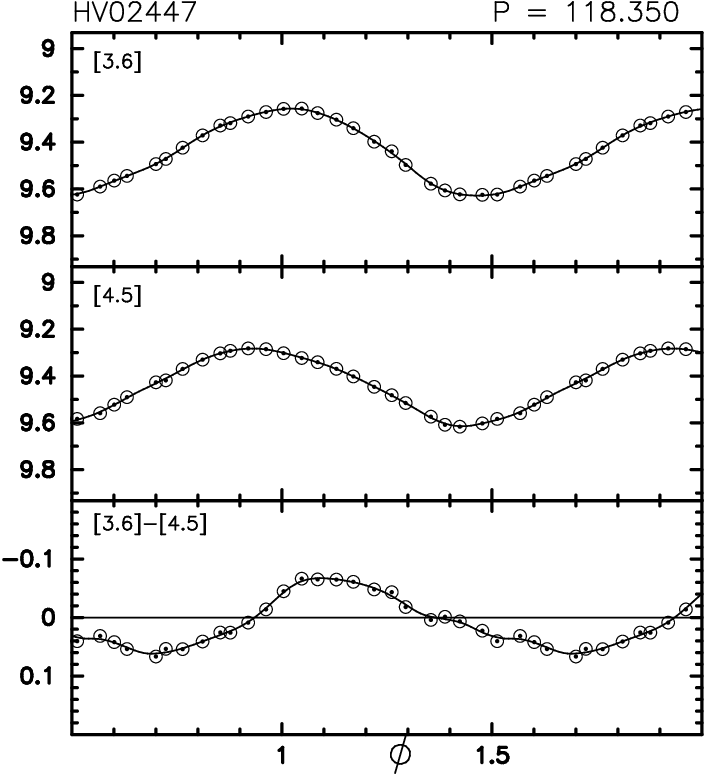} &
\includegraphics[width=50mm,angle=270]{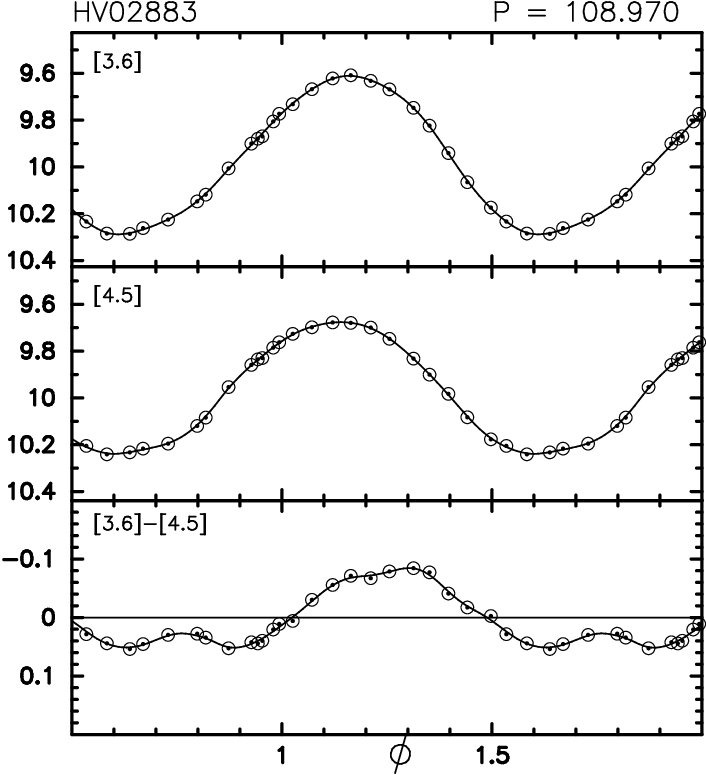} \\ 
\includegraphics[width=50mm,angle=270]{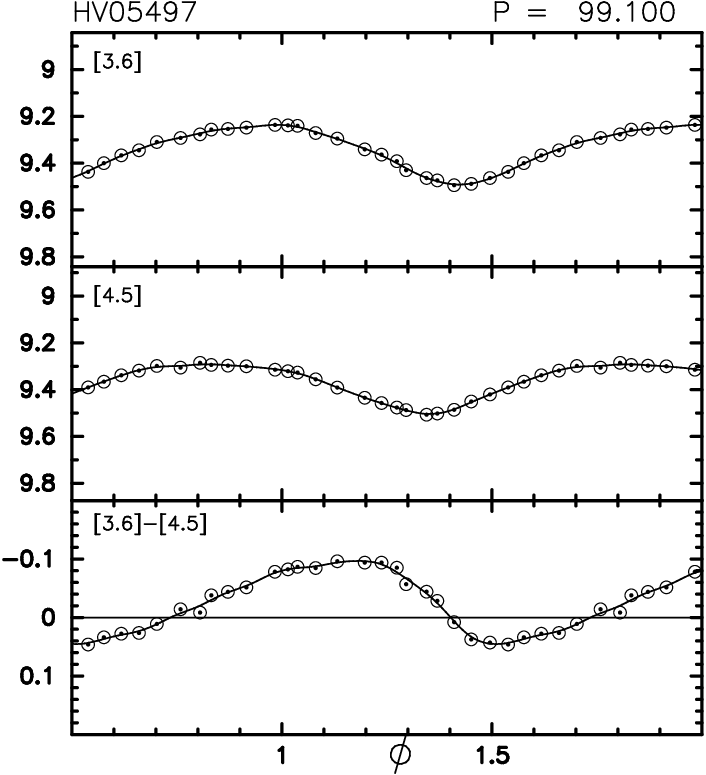} & 
\includegraphics[width=50mm,angle=270]{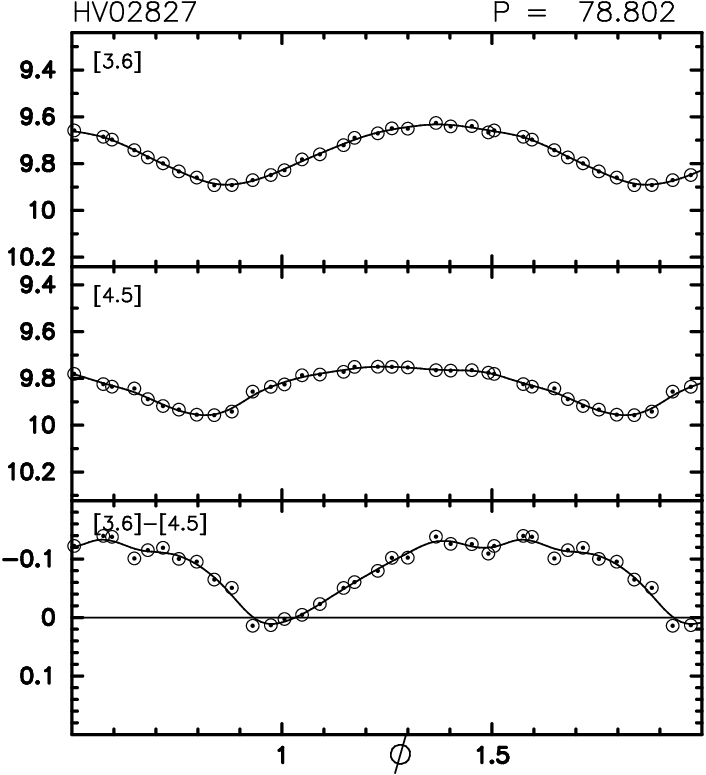} &
\includegraphics[width=50mm,angle=270]{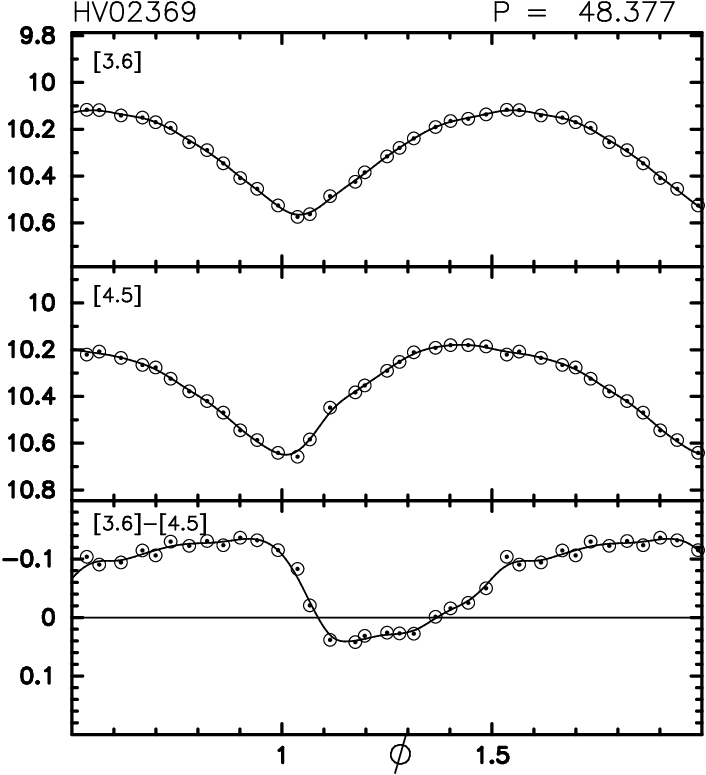} \\
\includegraphics[width=50mm,angle=270]{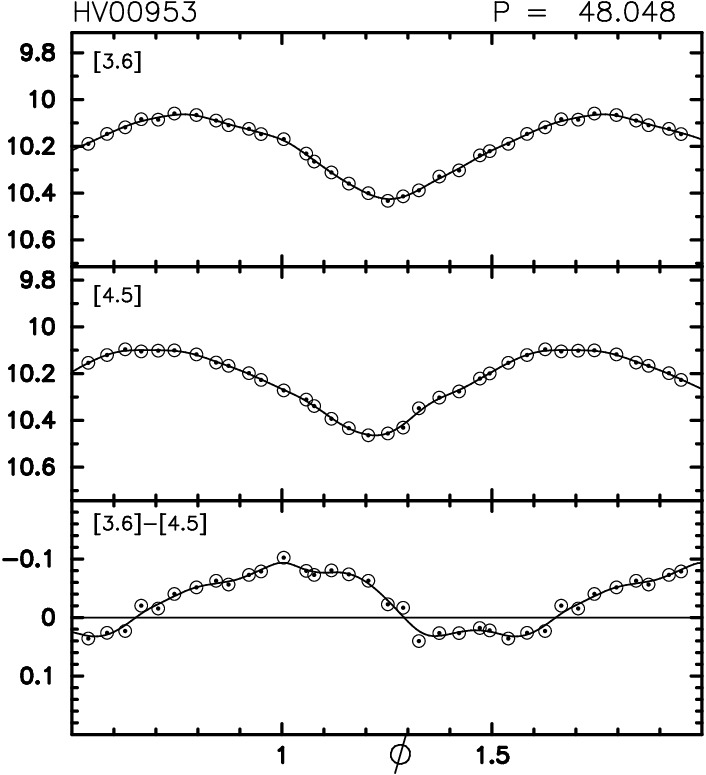} & 
\includegraphics[width=50mm,angle=270]{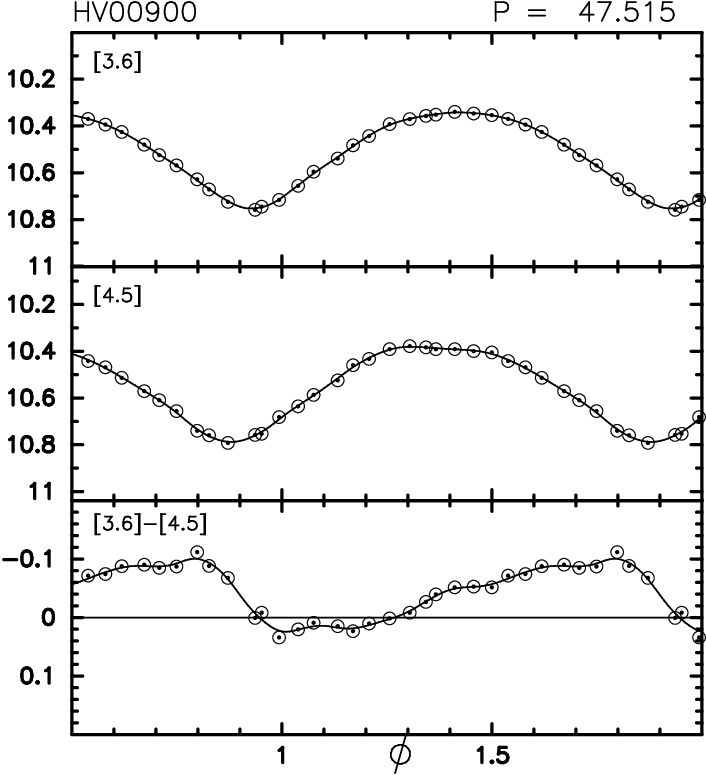} & 
\includegraphics[width=50mm,angle=270]{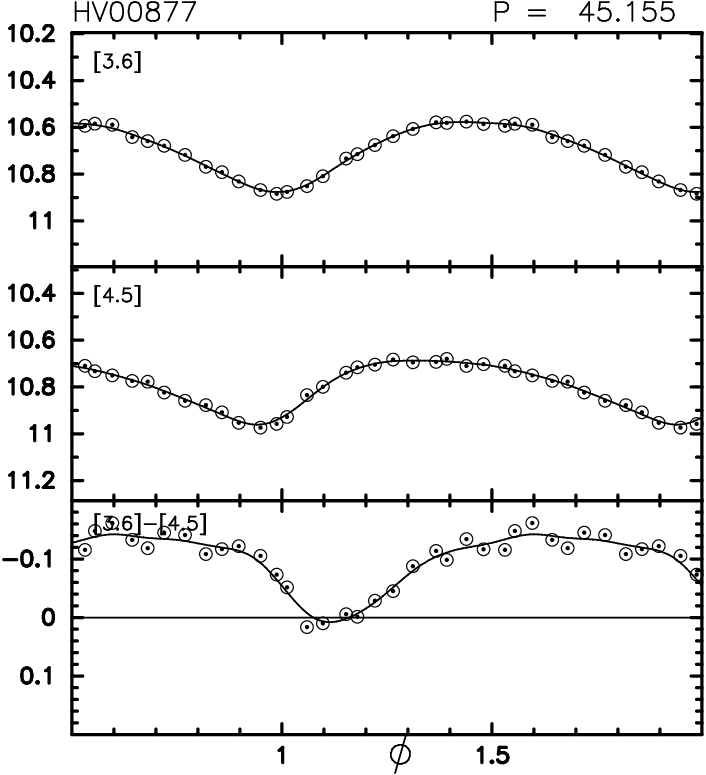} \\
\end{array}$ 
\end{center} 
\end{figure}
\begin{figure} 
 \begin{center}$ 
 \begin{array}{ccc} 
\includegraphics[width=50mm,angle=270]{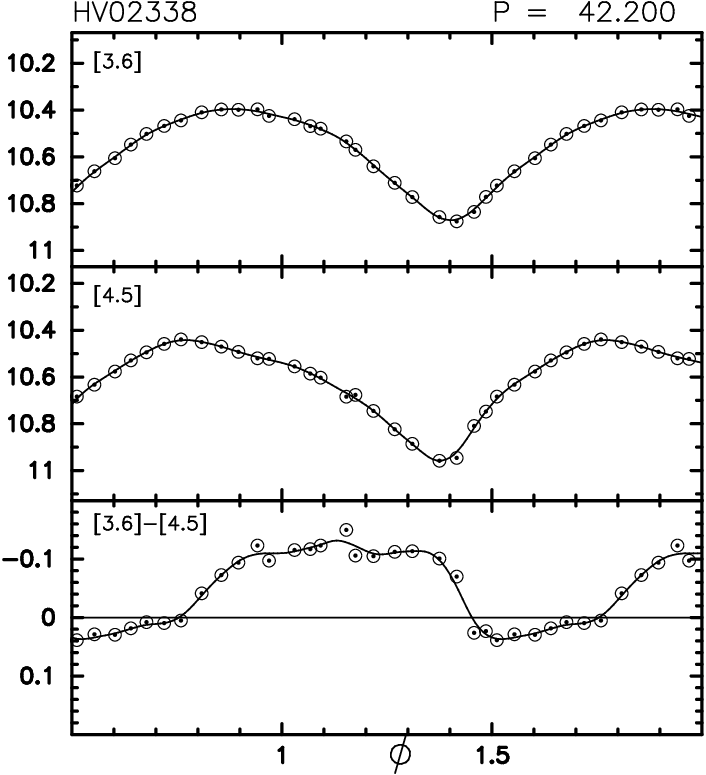} & 
 \includegraphics[width=50mm,angle=270]{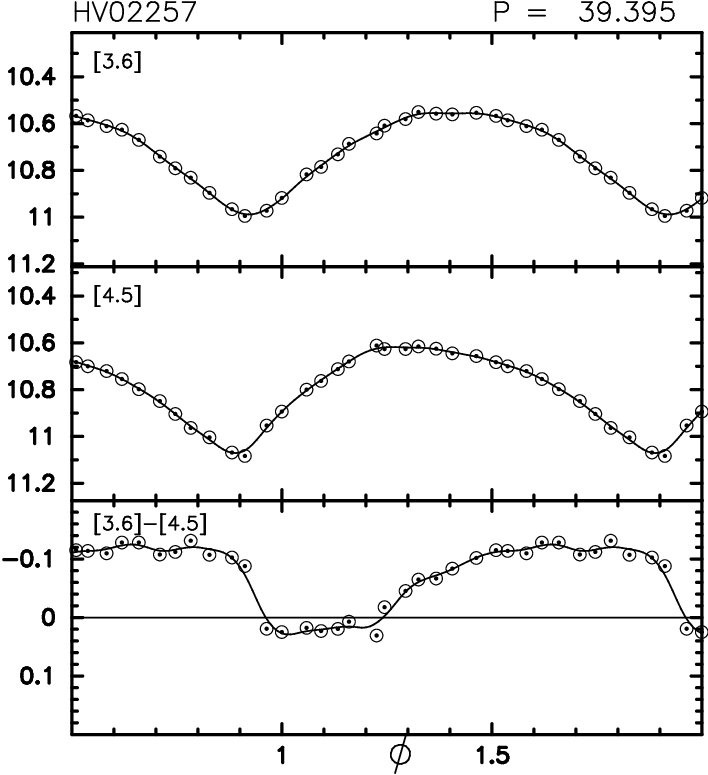} & 
\includegraphics[width=50mm,angle=270]{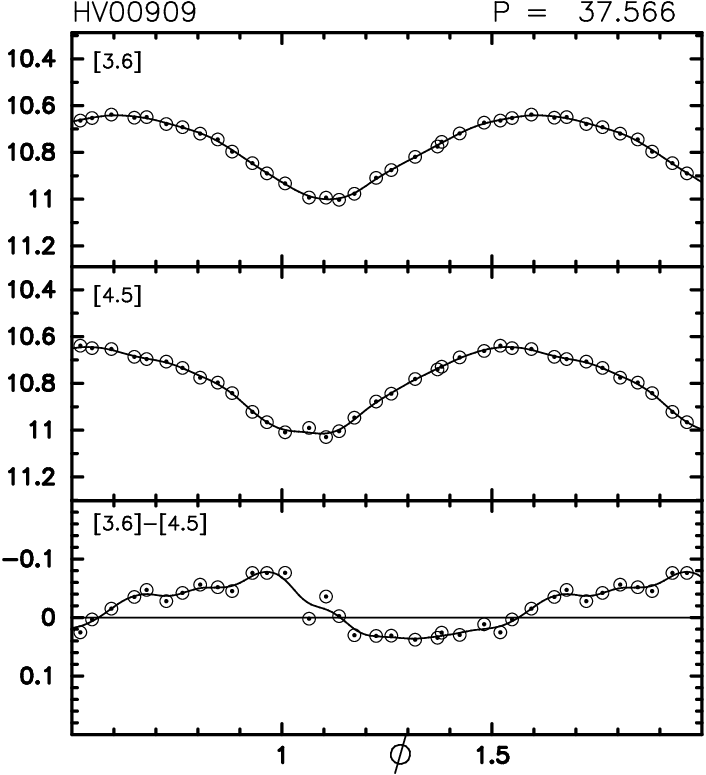} \\ 
\includegraphics[width=50mm,angle=270]{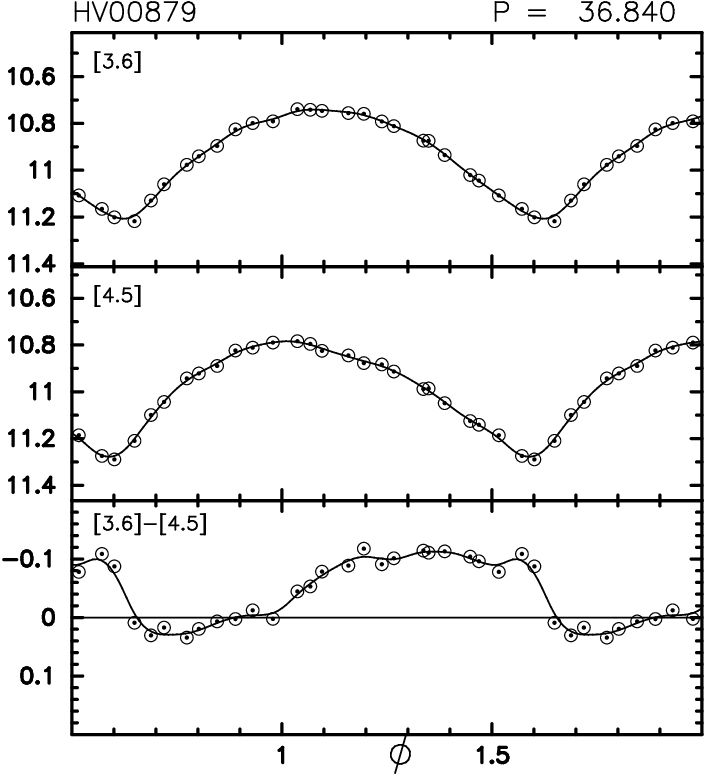} & 
\includegraphics[width=50mm,angle=270]{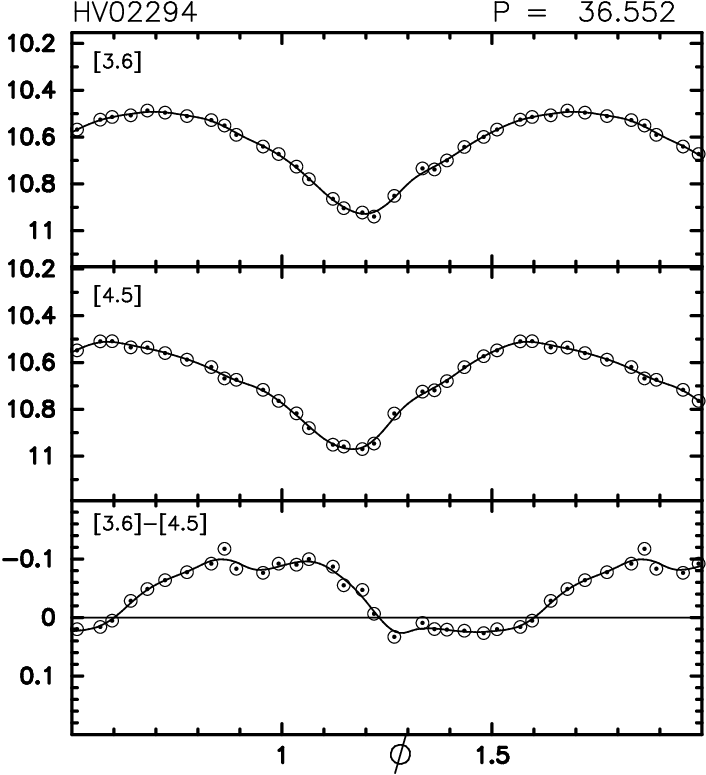} &
\includegraphics[width=50mm,angle=270]{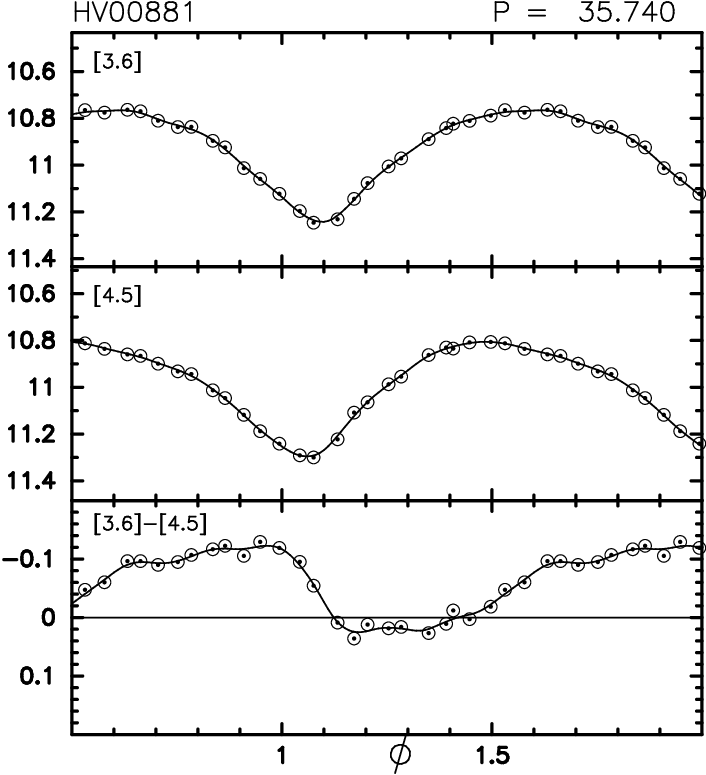} \\
\includegraphics[width=50mm,angle=270]{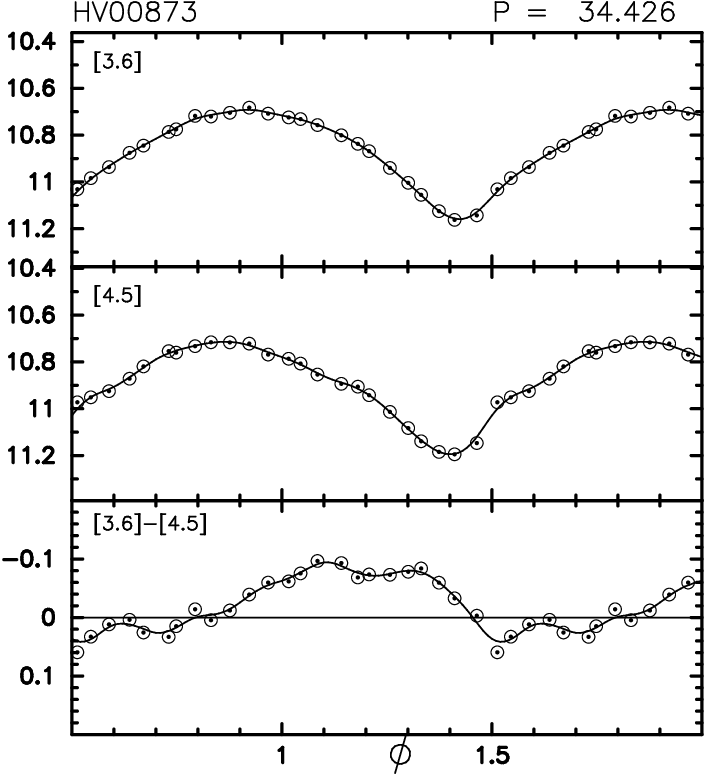} & 
\includegraphics[width=50mm,angle=270]{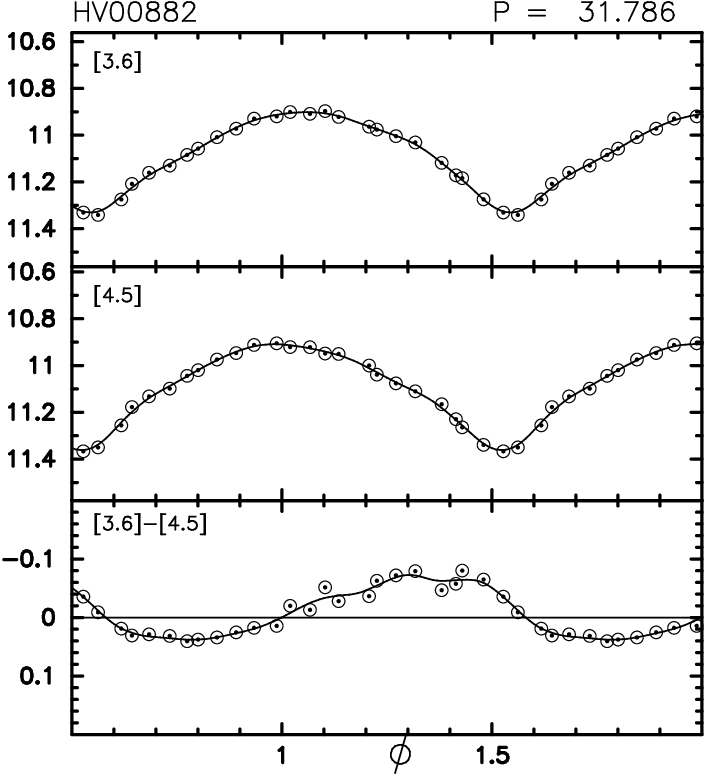} &
\includegraphics[width=50mm,angle=270]{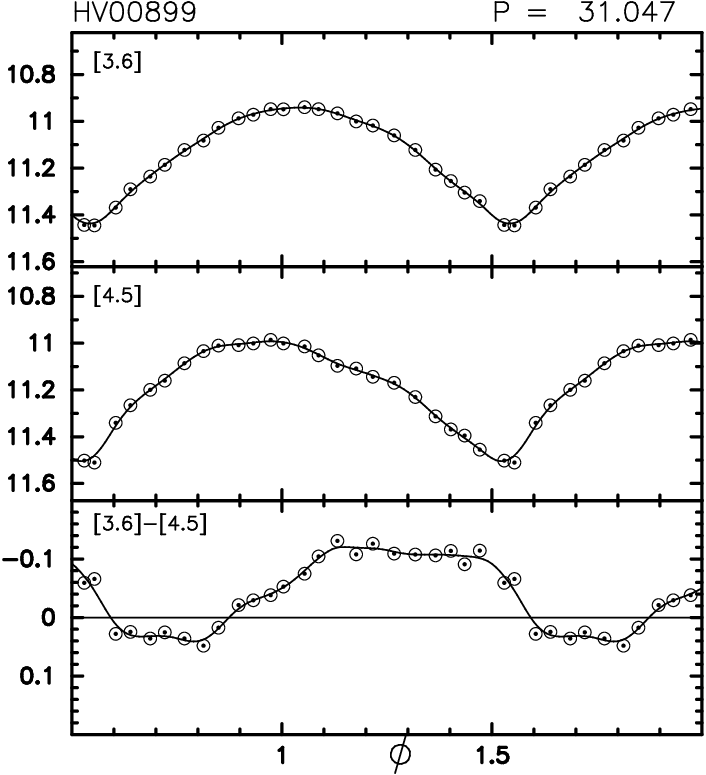} \\
\end{array}$ 
\end{center} 
\end{figure}
\begin{figure} 
 \begin{center}$ 
 \begin{array}{ccc} 
\includegraphics[width=50mm,angle=270]{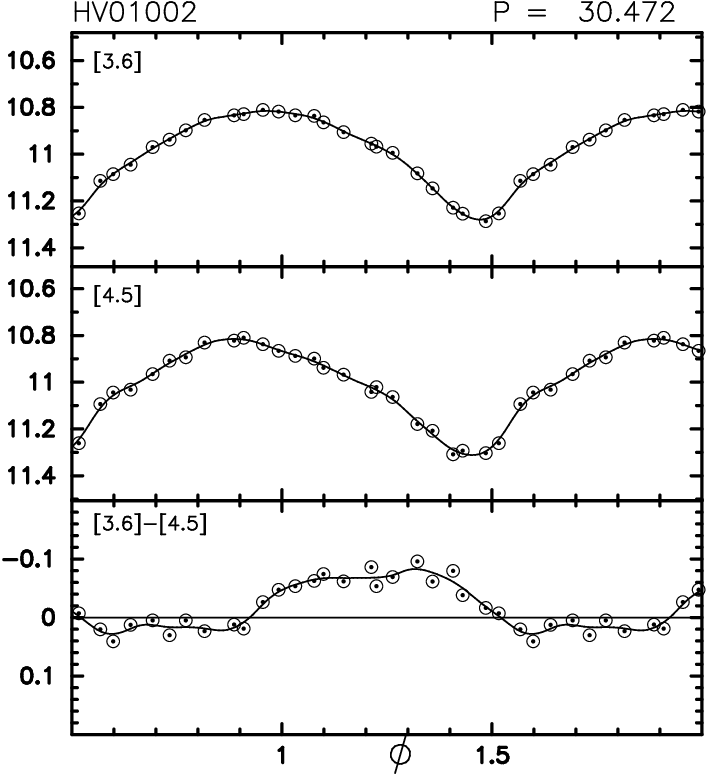} & 
\includegraphics[width=50mm,angle=270]{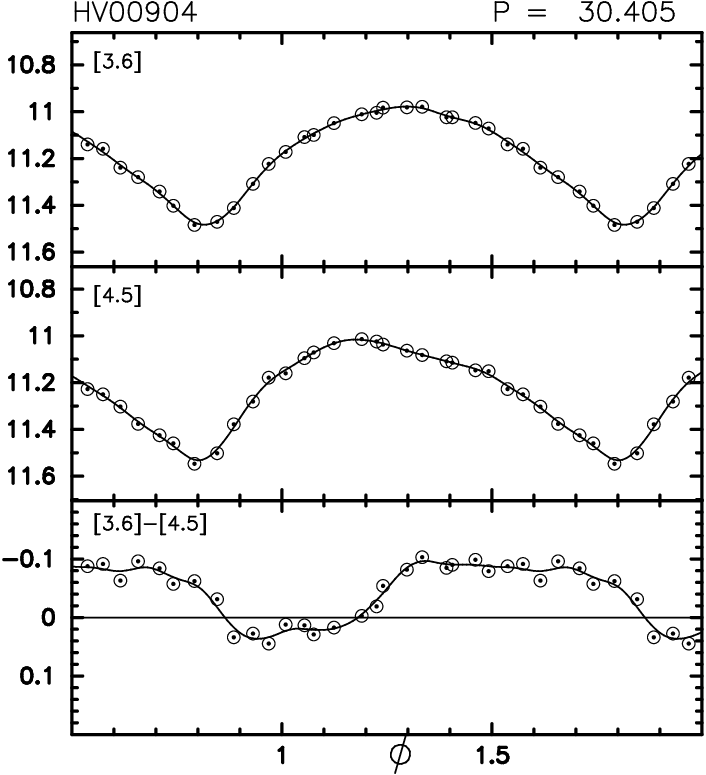} & 
\includegraphics[width=50mm,angle=270]{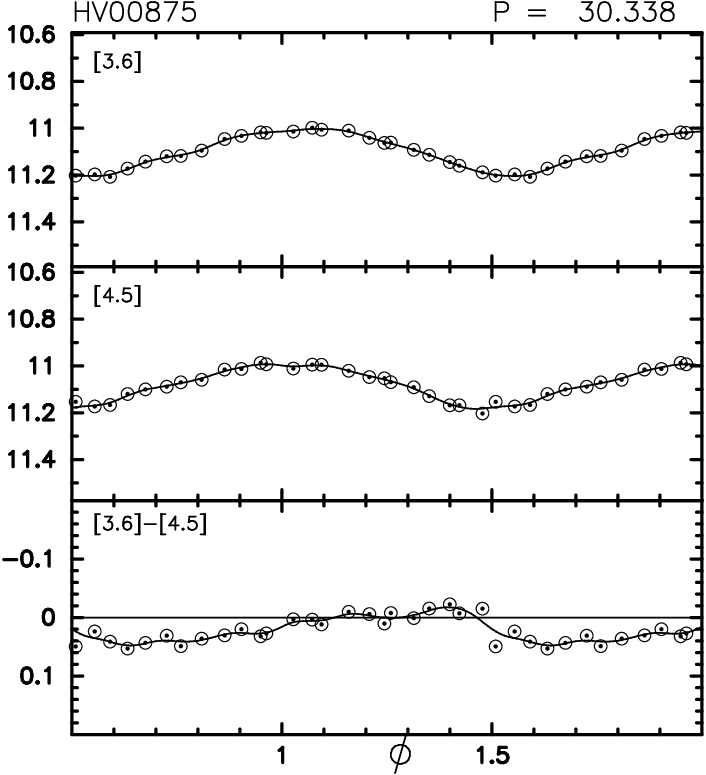} \\ 
\includegraphics[width=50mm,angle=270]{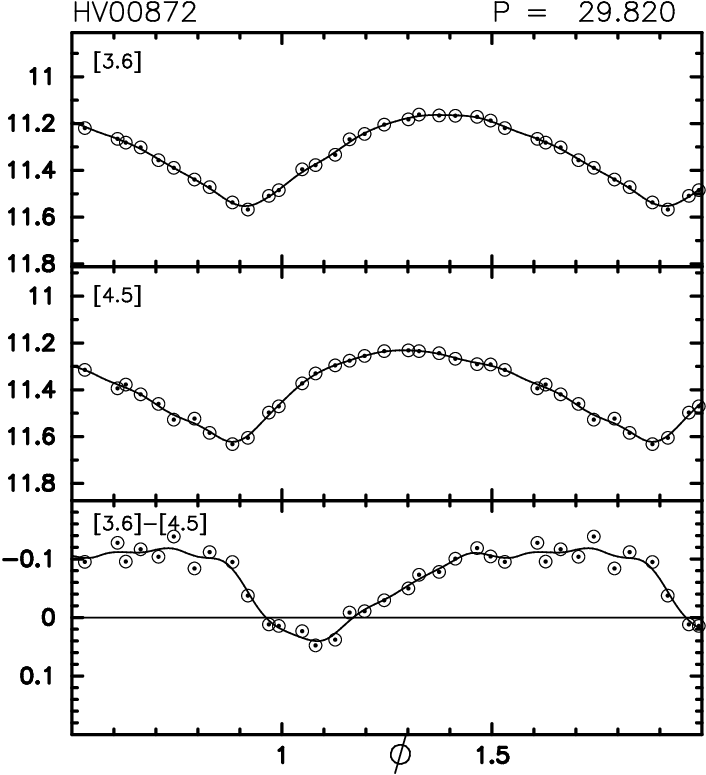} & 
\includegraphics[width=50mm,angle=270]{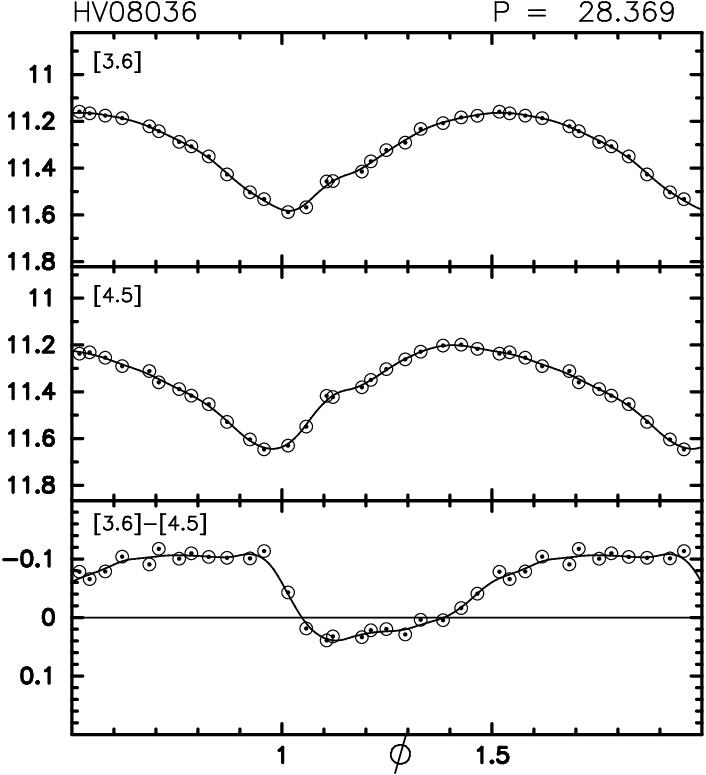} &
\includegraphics[width=50mm,angle=270]{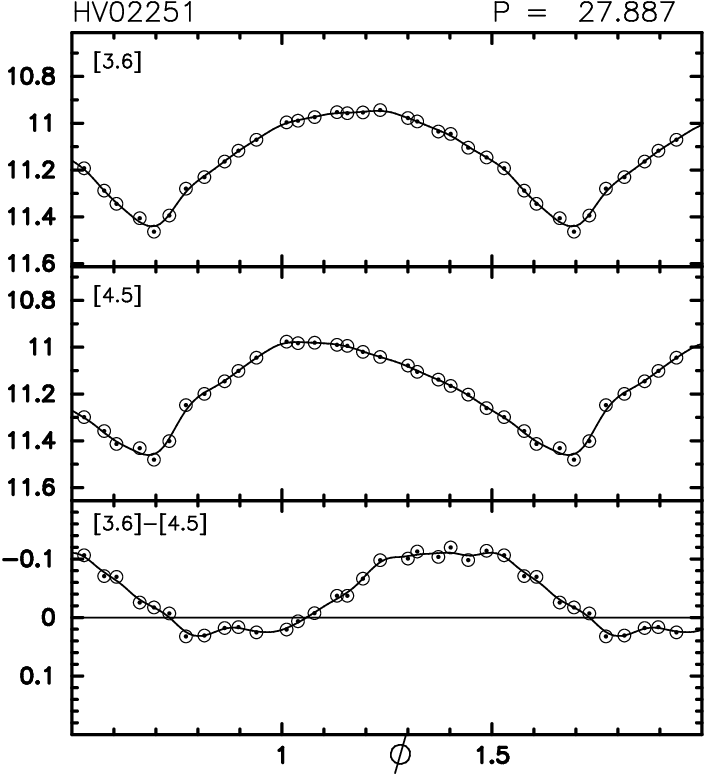} \\ 
\includegraphics[width=50mm,angle=270]{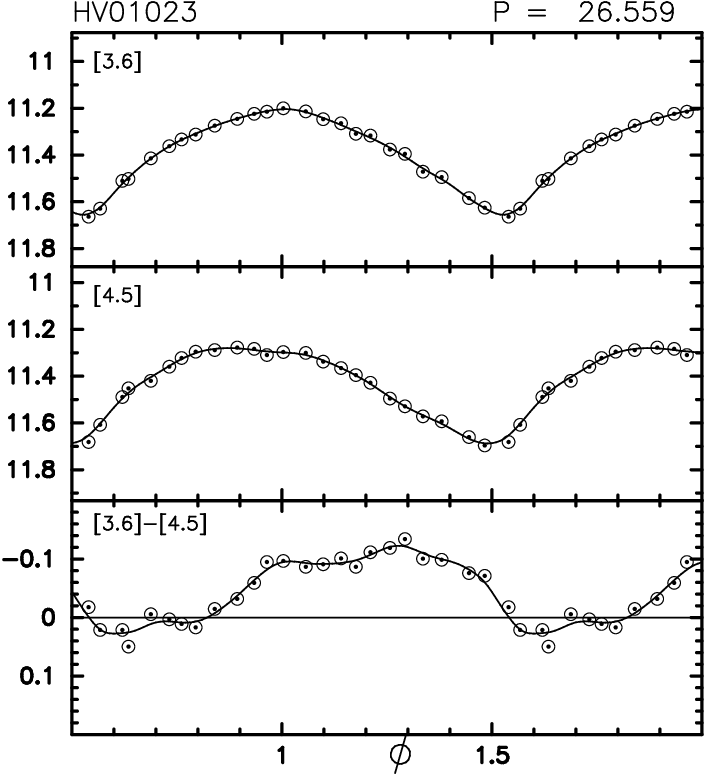} & 
\includegraphics[width=50mm,angle=270]{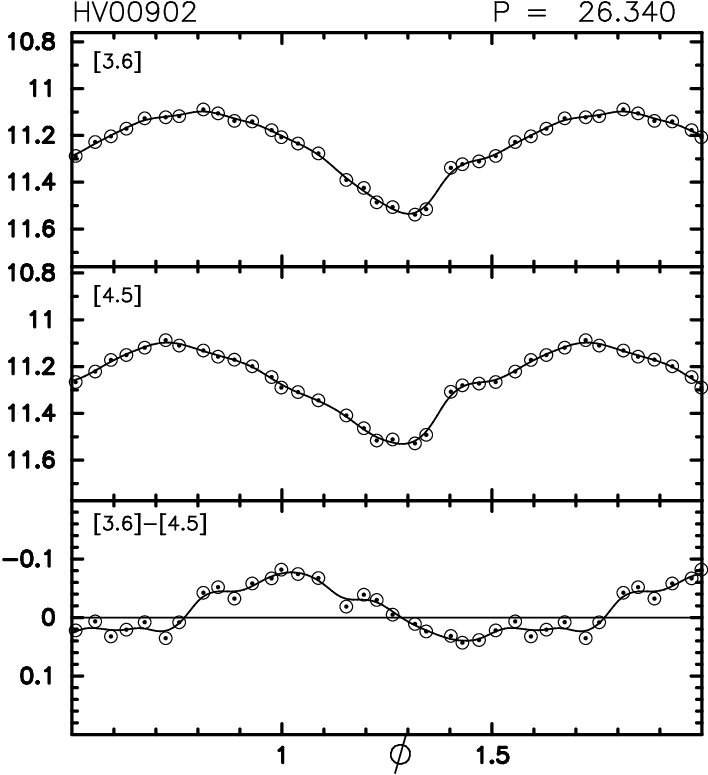} & 
\includegraphics[width=50mm,angle=270]{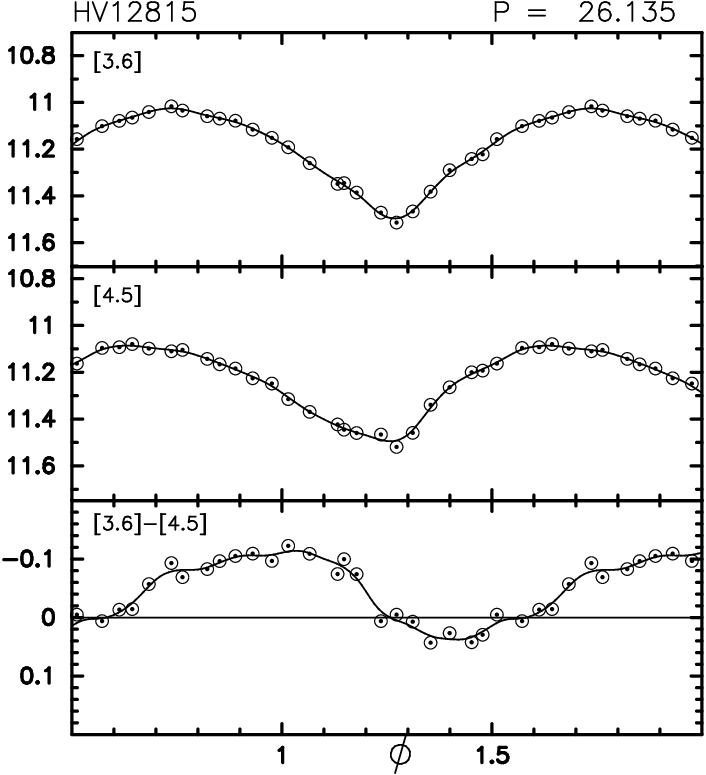} \\
\end{array}$ 
\end{center} 
\end{figure}
\begin{figure} 
 \begin{center}$ 
 \begin{array}{ccc} 
\includegraphics[width=50mm,angle=270]{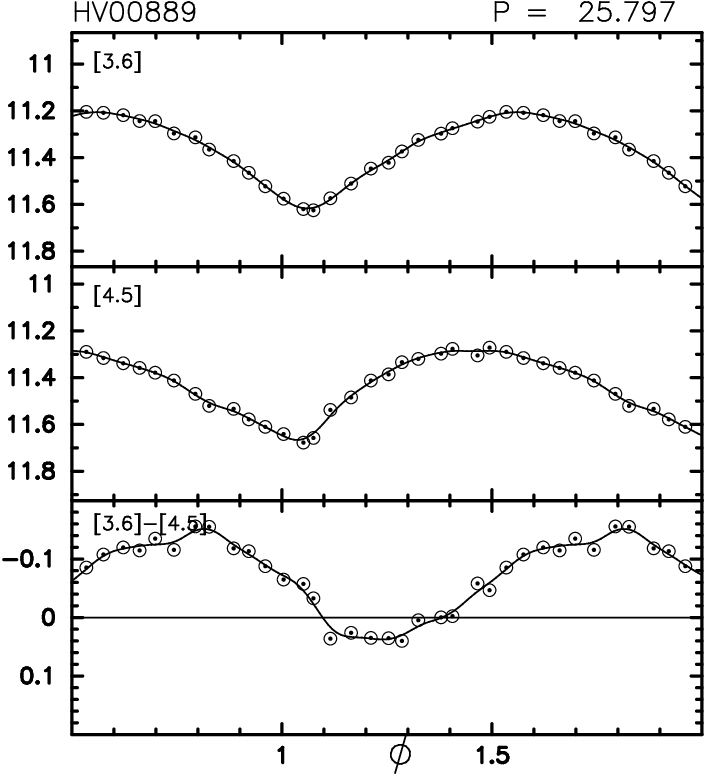} & 
\includegraphics[width=50mm,angle=270]{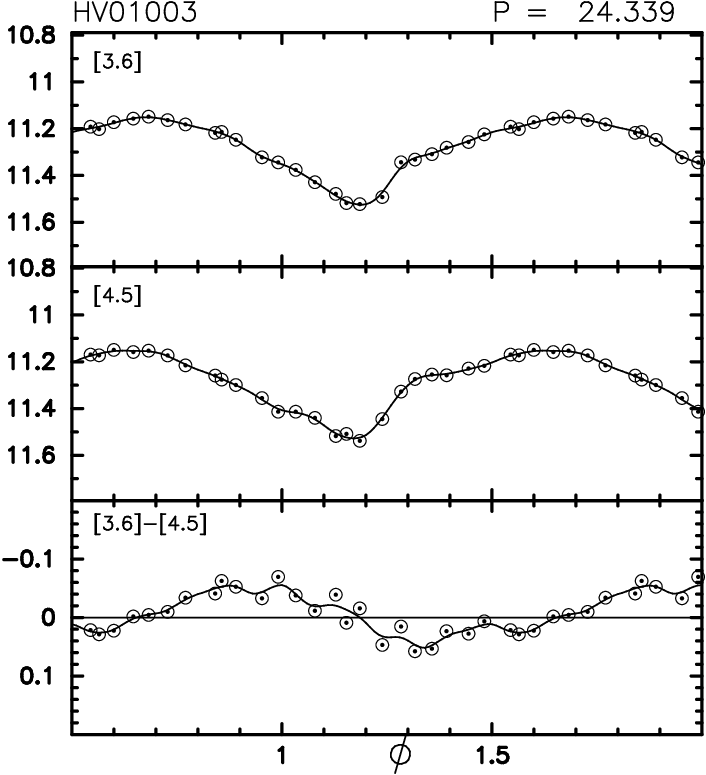} &
\includegraphics[width=50mm,angle=270]{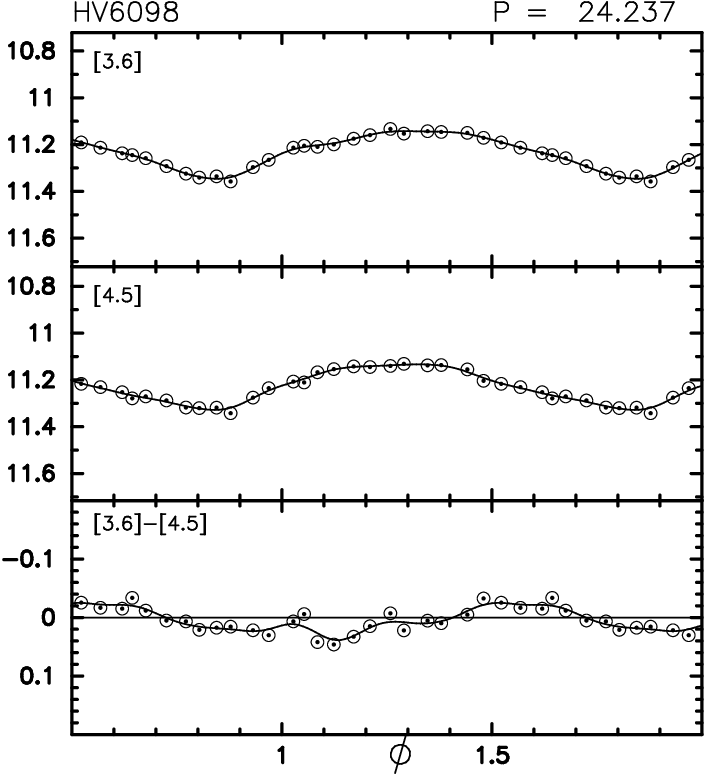} \\
\includegraphics[width=50mm,angle=270]{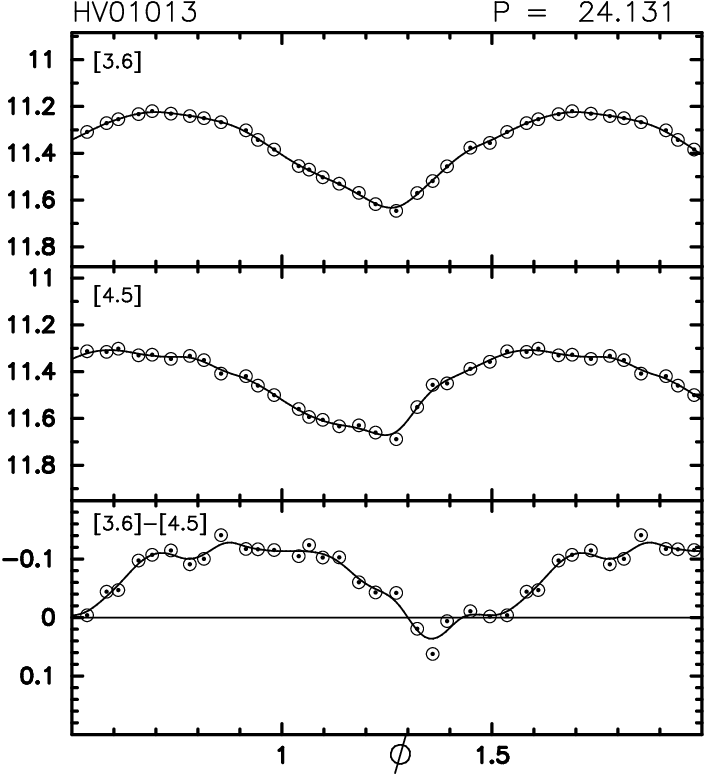} & 
\includegraphics[width=50mm,angle=270]{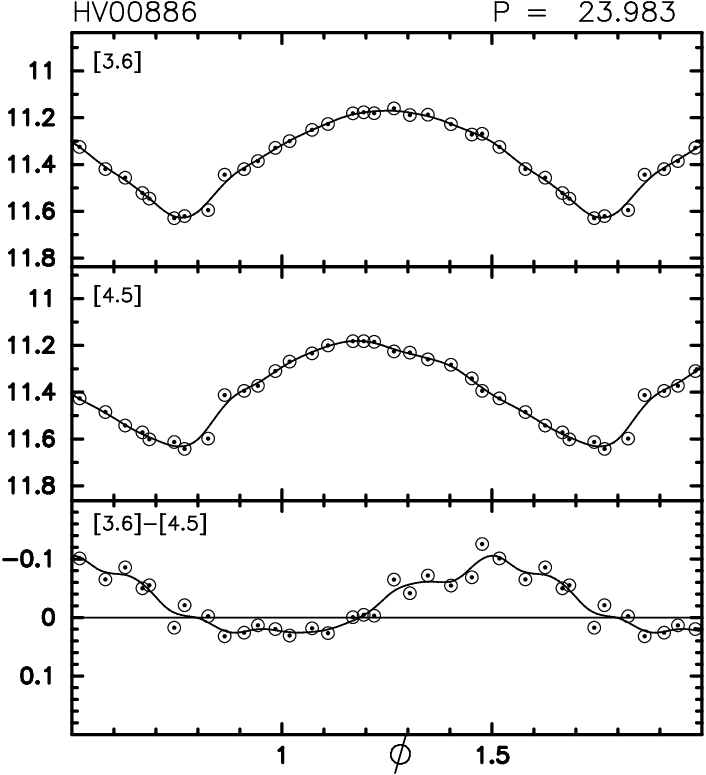} &
\includegraphics[width=50mm,angle=270]{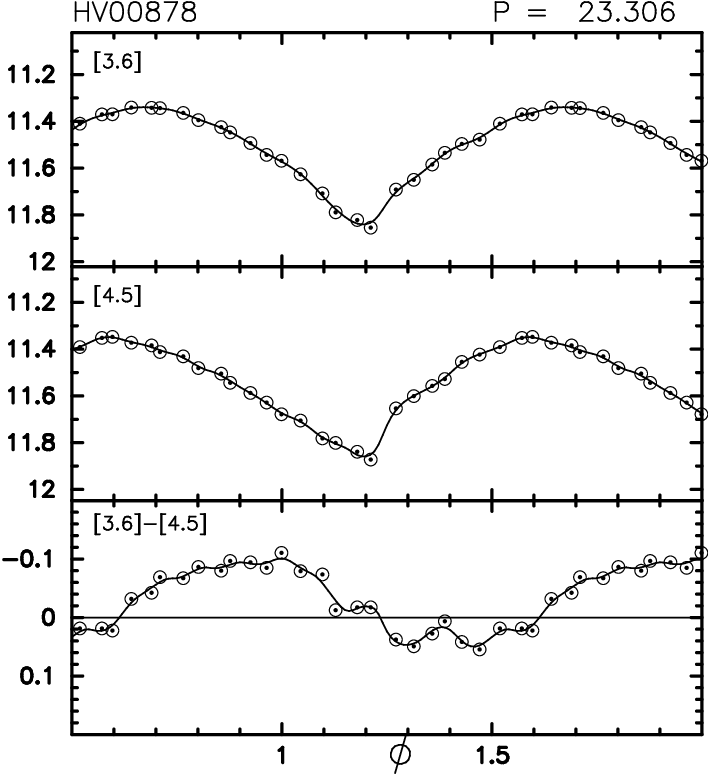} \\
\includegraphics[width=50mm,angle=270]{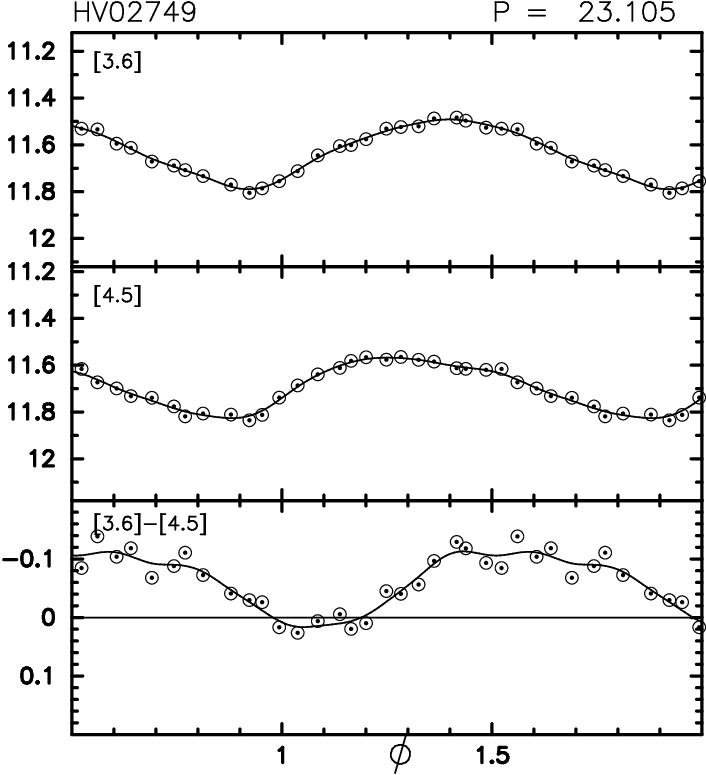} & 
\includegraphics[width=50mm,angle=270]{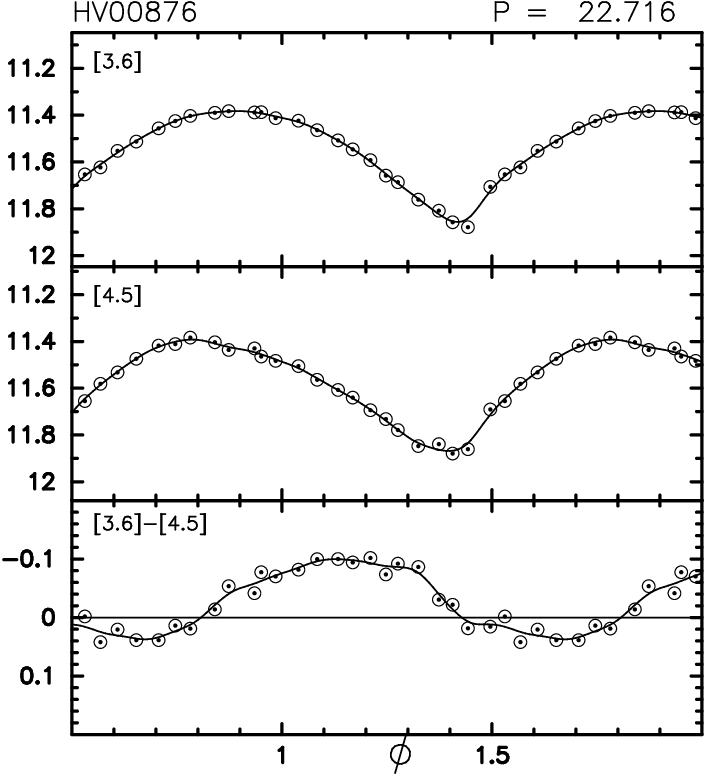} &
\includegraphics[width=50mm,angle=270]{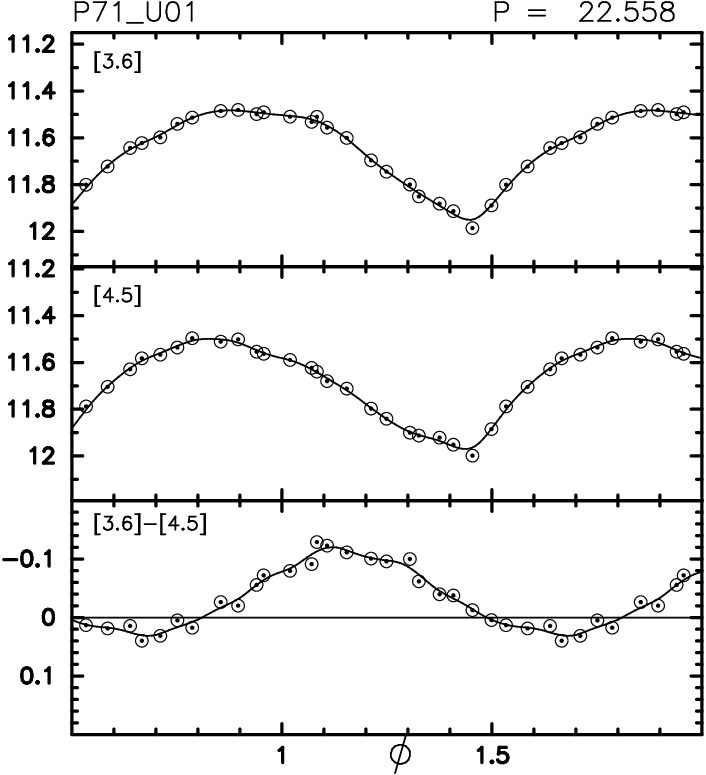} \\
\end{array}$ 
\end{center} 
\end{figure}
\begin{figure} 
 \begin{center}$ 
 \begin{array}{ccc}  
\includegraphics[width=50mm,angle=270]{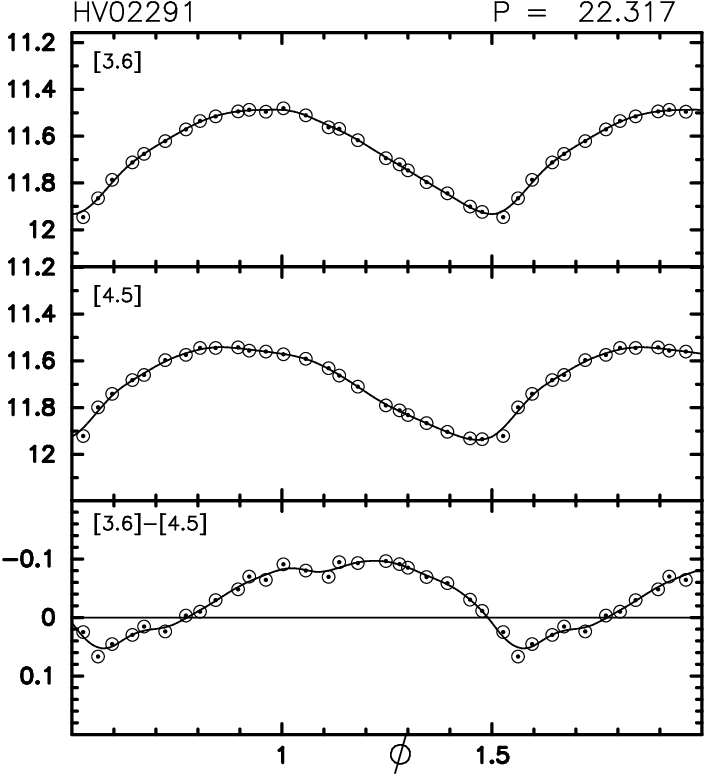} & 
\includegraphics[width=50mm,angle=270]{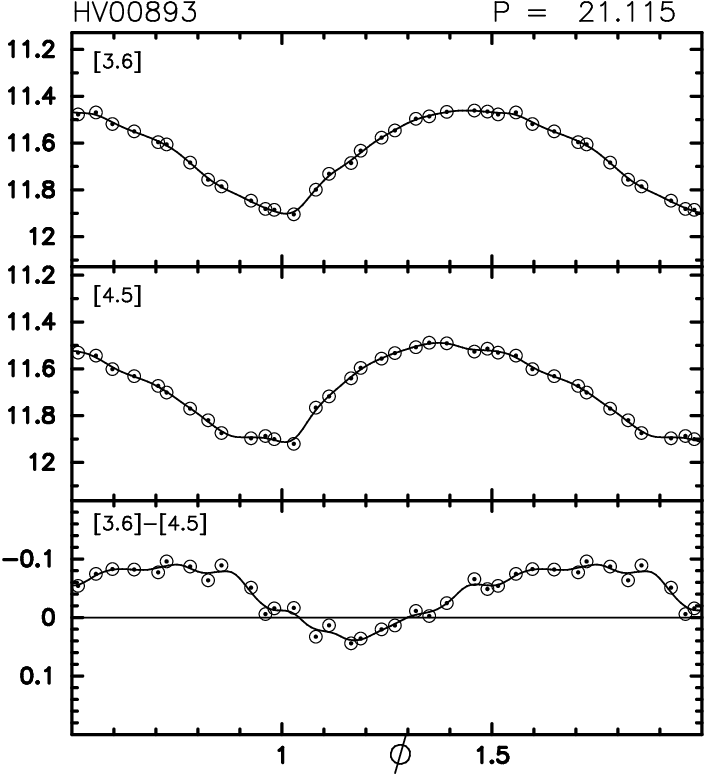} & 
\includegraphics[width=50mm,angle=270]{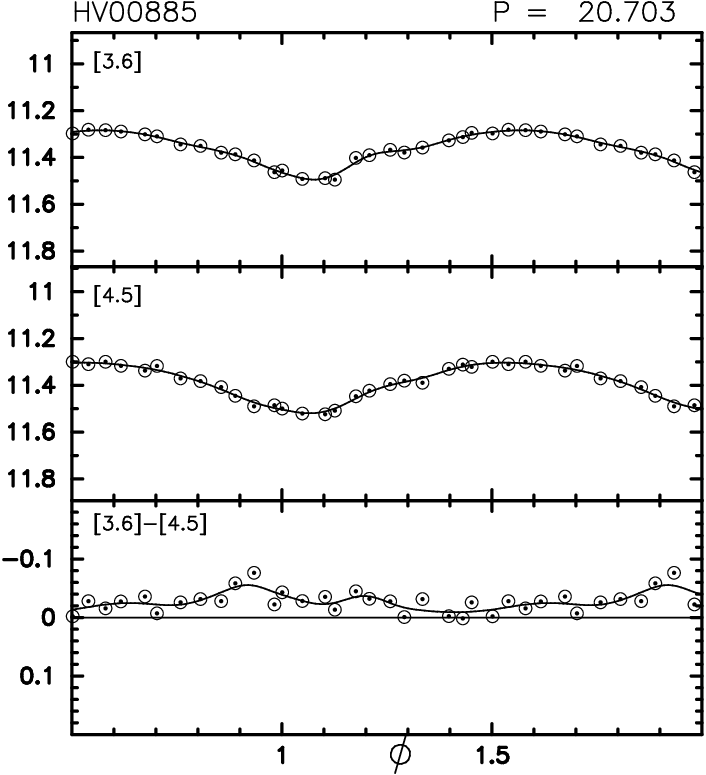} \\ 
\includegraphics[width=50mm,angle=270]{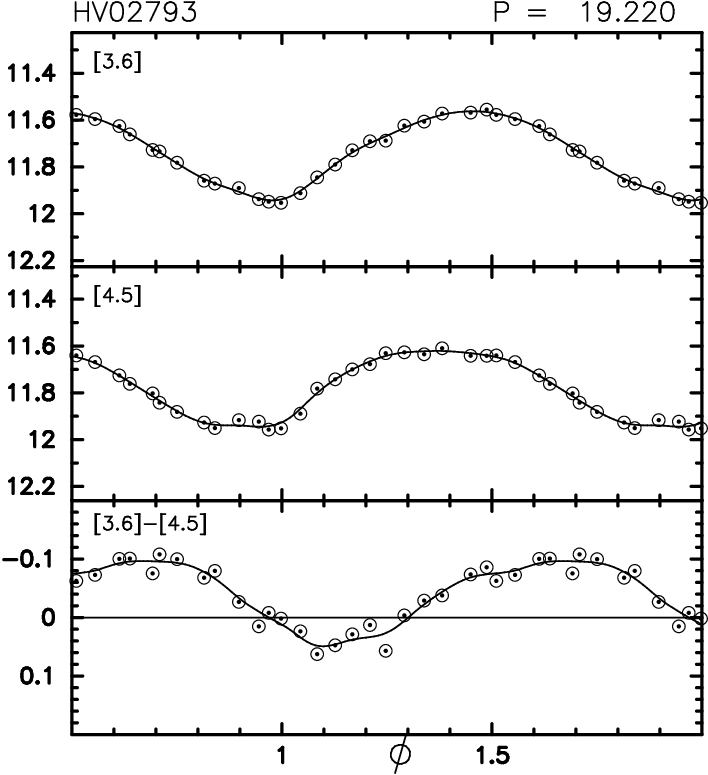} & 
\includegraphics[width=50mm,angle=270]{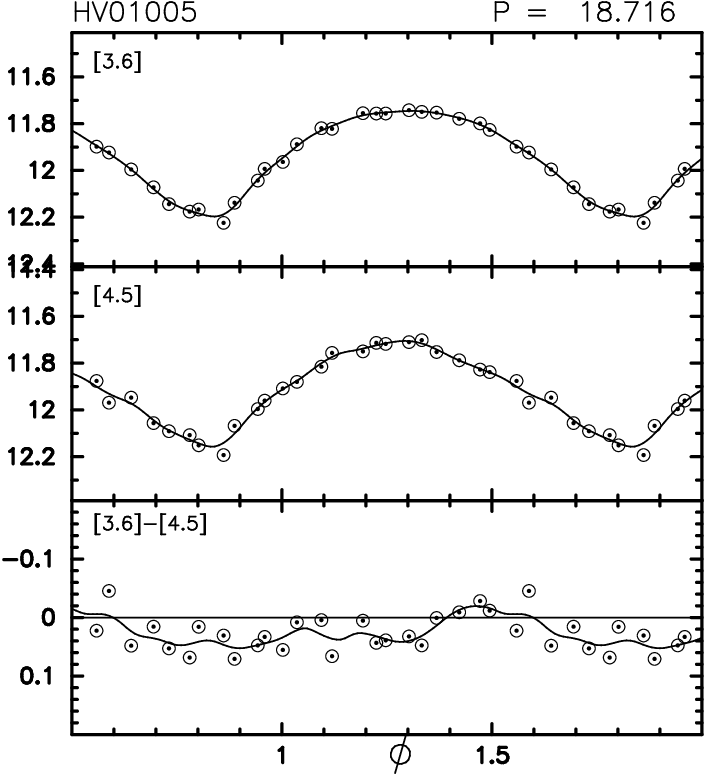} &
\includegraphics[width=50mm,angle=270]{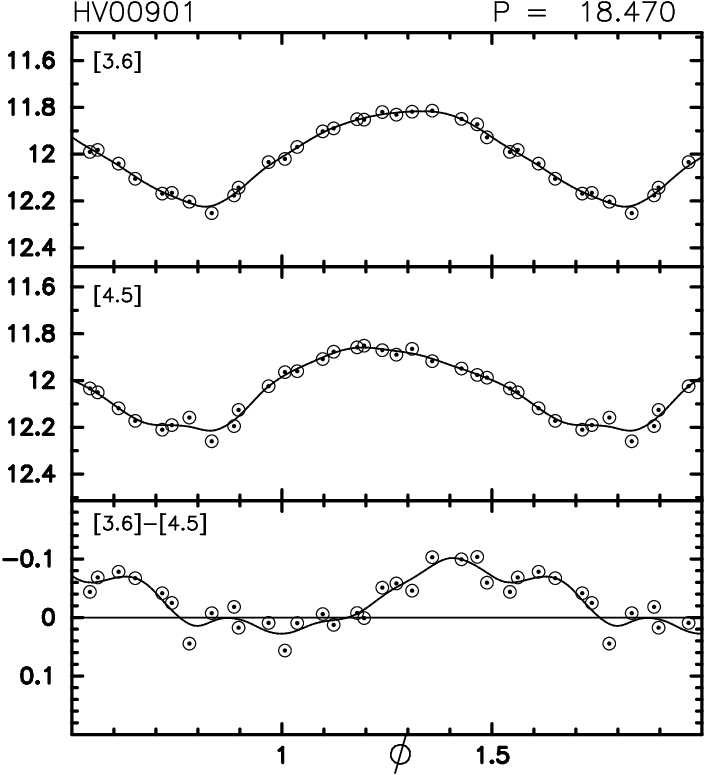} \\
\includegraphics[width=50mm,angle=270]{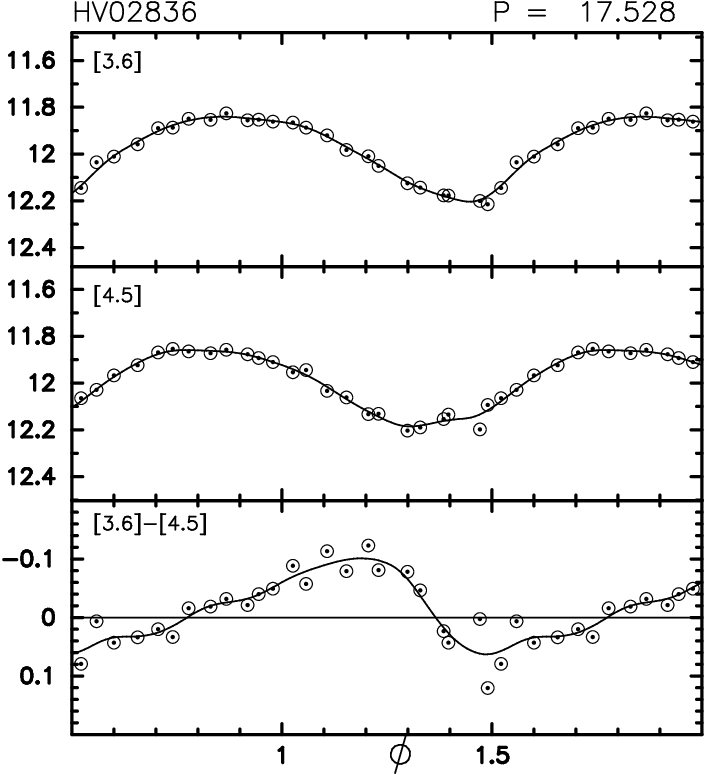} & 
\includegraphics[width=50mm,angle=270]{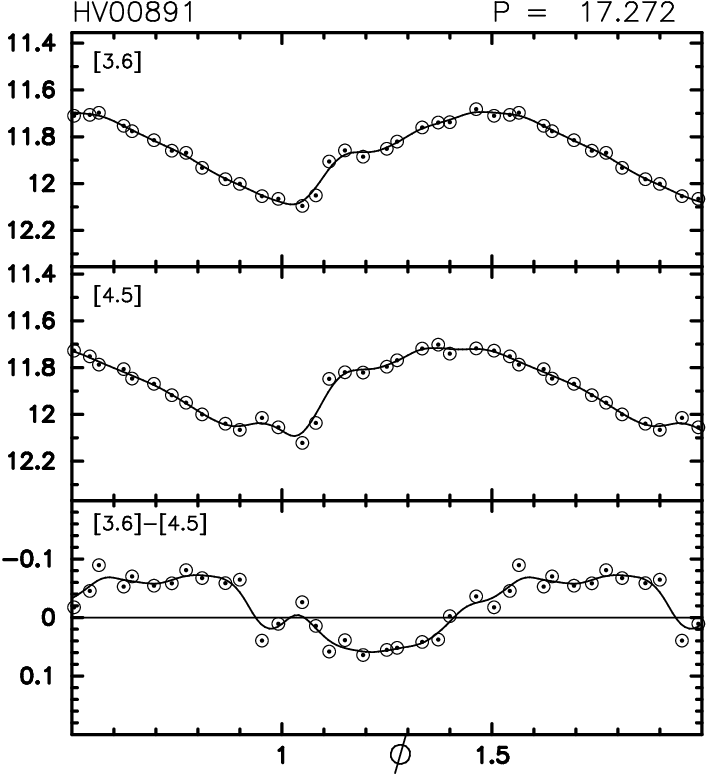} &
\includegraphics[width=50mm,angle=270]{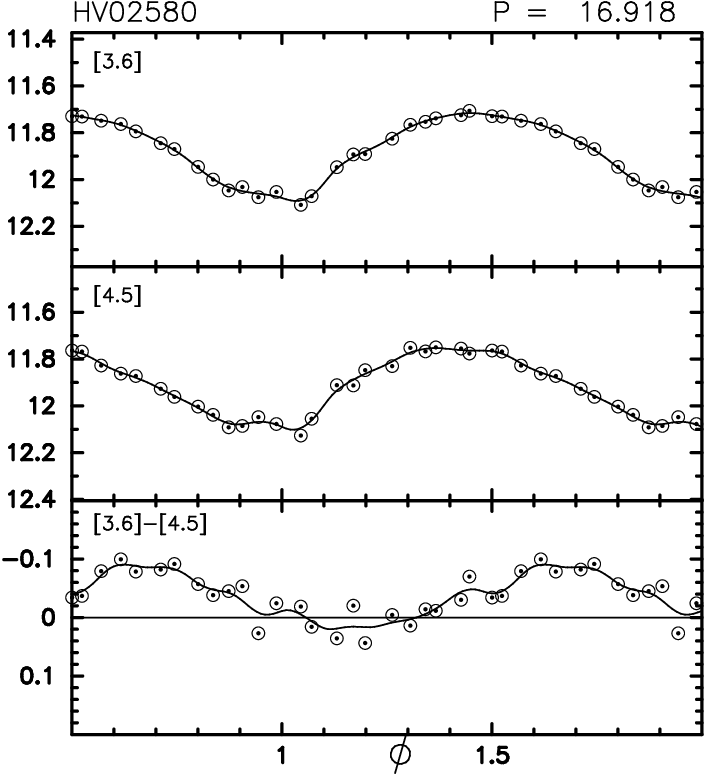} \\
\end{array}$ 
\end{center} 
\end{figure}
\begin{figure} 
 \begin{center}$ 
 \begin{array}{ccc} 
\includegraphics[width=50mm,angle=270]{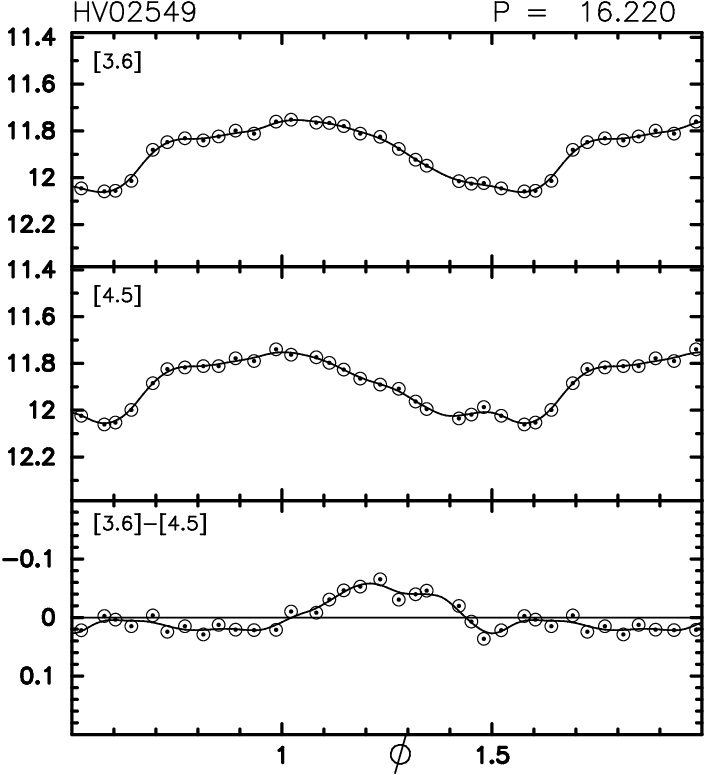} & 
\includegraphics[width=50mm,angle=270]{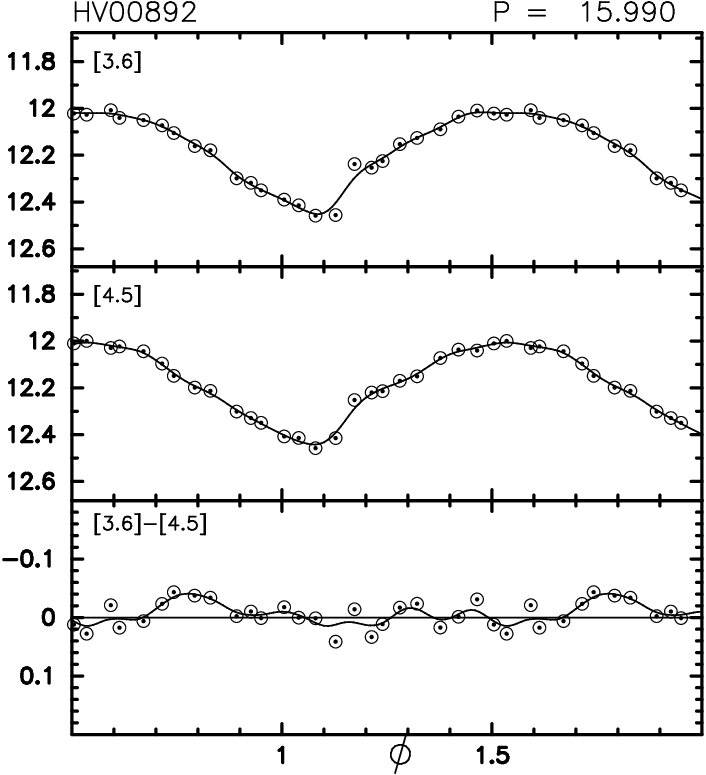} & 
\includegraphics[width=50mm,angle=270]{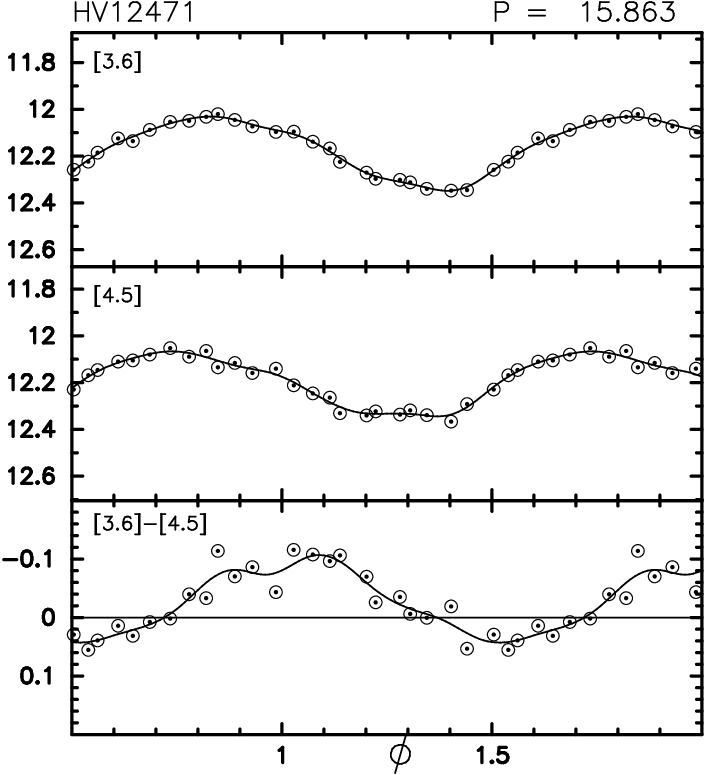} \\ 
\includegraphics[width=50mm,angle=270]{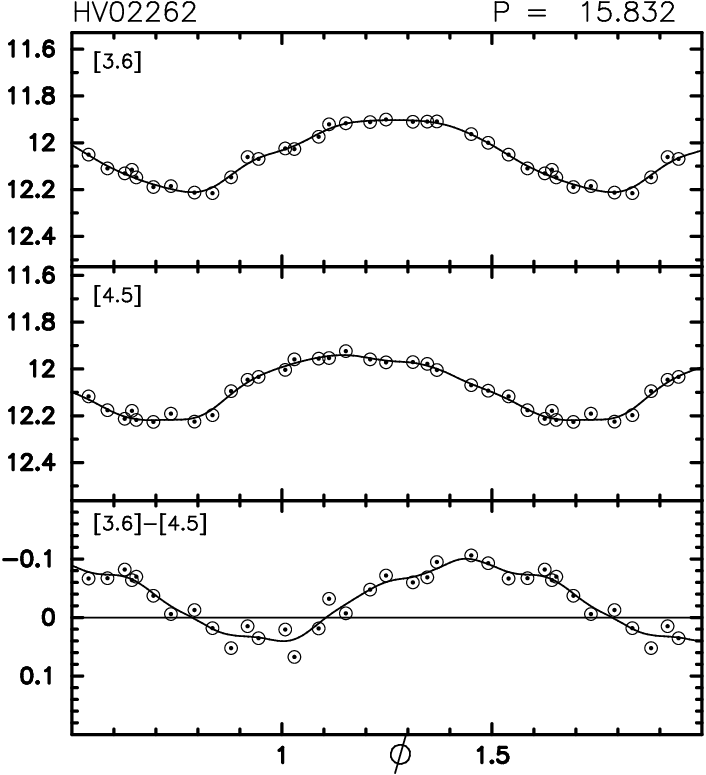} & 
\includegraphics[width=50mm,angle=270]{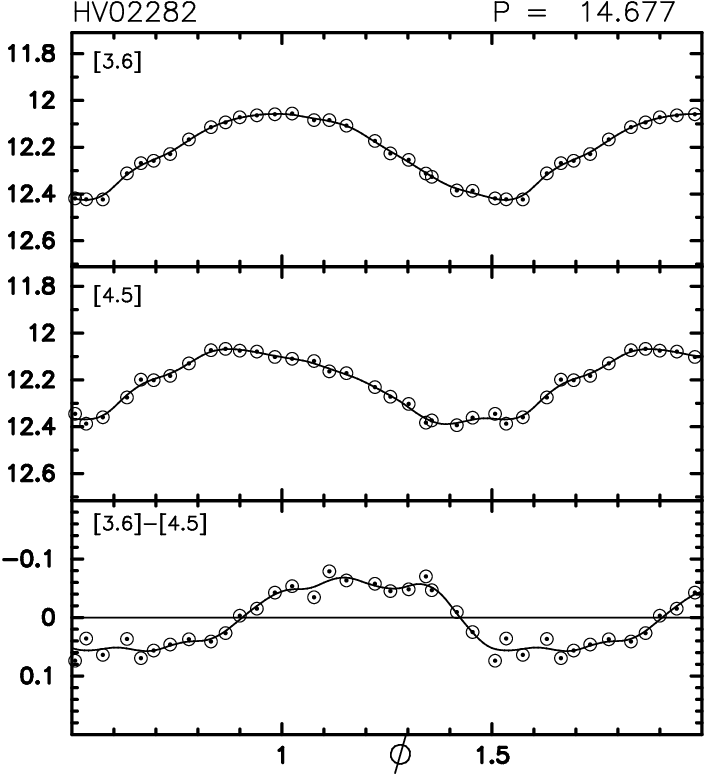} & 
\includegraphics[width=50mm,angle=270]{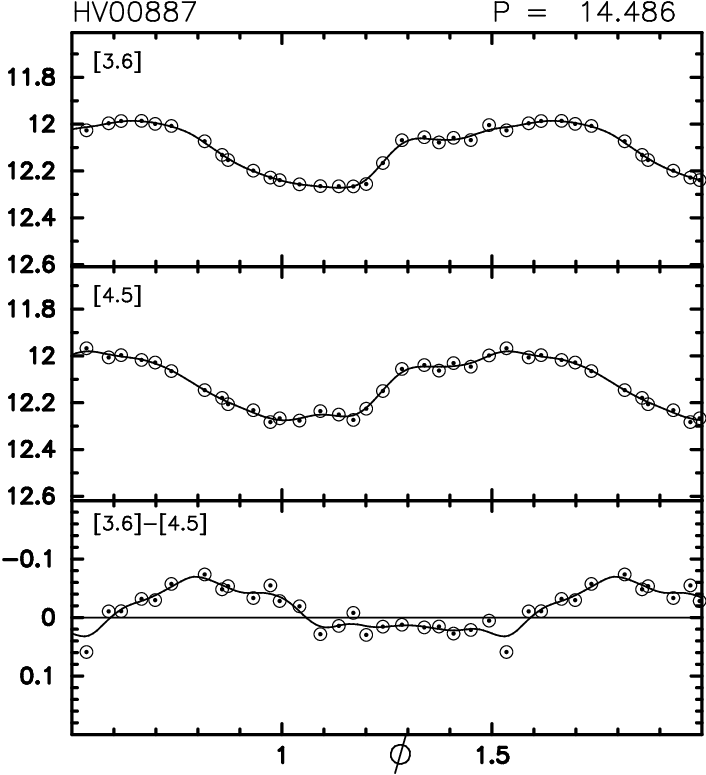} \\
\includegraphics[width=50mm,angle=270]{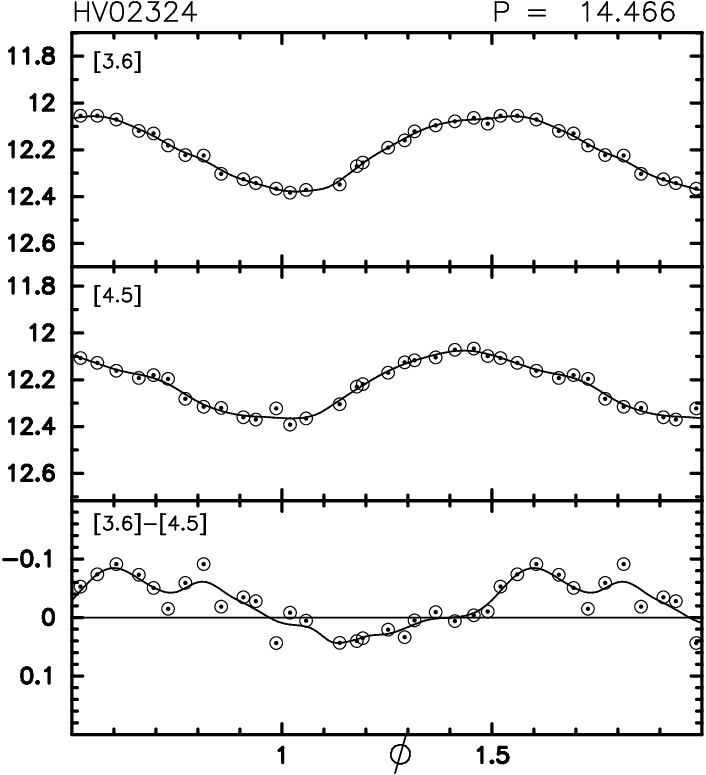} & 
\includegraphics[width=50mm,angle=270]{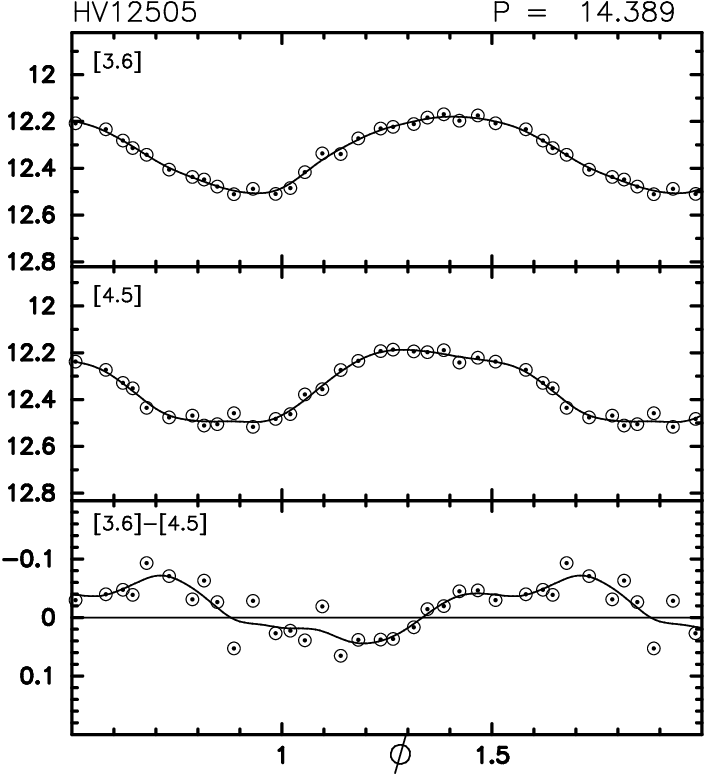} & 
\includegraphics[width=50mm,angle=270]{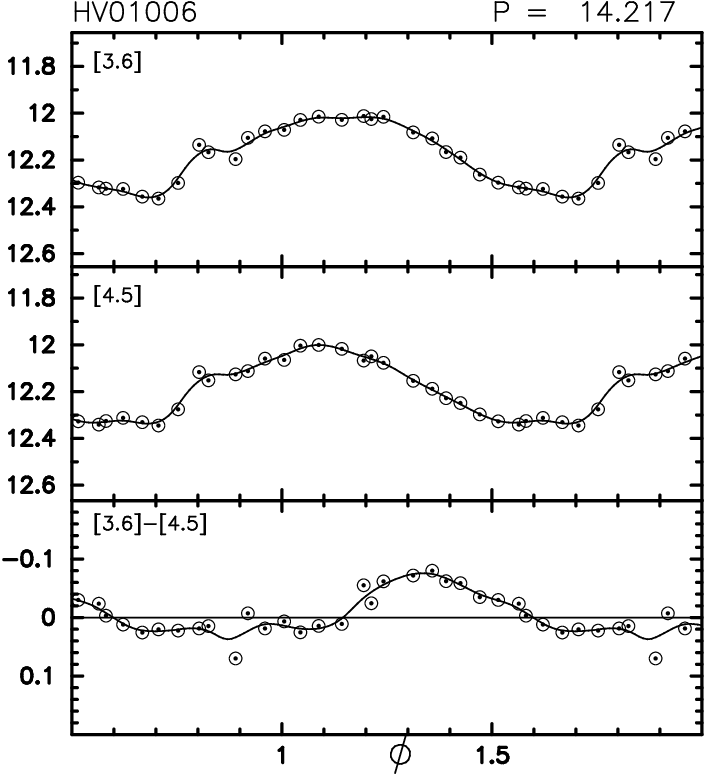} \\
\end{array}$ 
\end{center} 
\end{figure}
\begin{figure} 
 \begin{center}$ 
 \begin{array}{ccc} 
\includegraphics[width=50mm,angle=270]{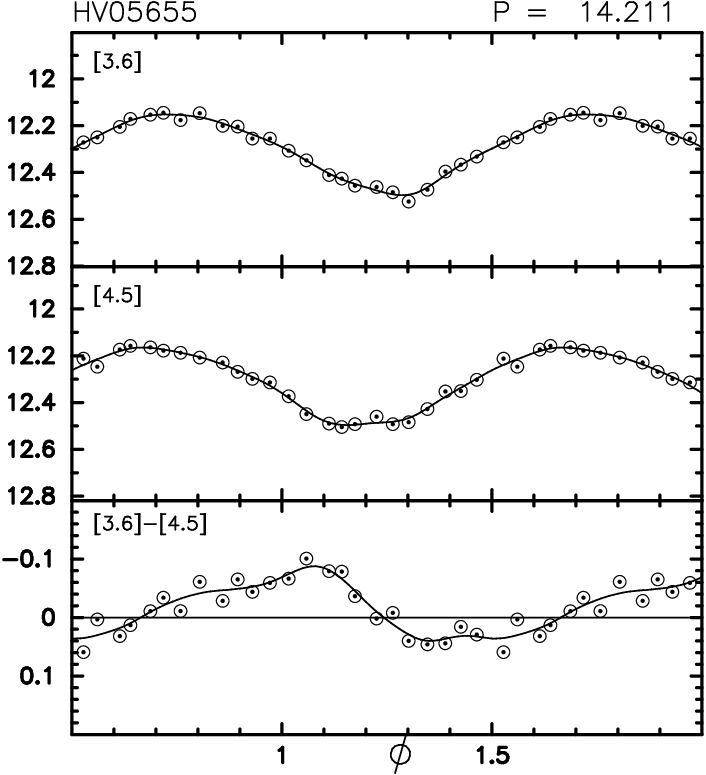} & 
\includegraphics[width=50mm,angle=270]{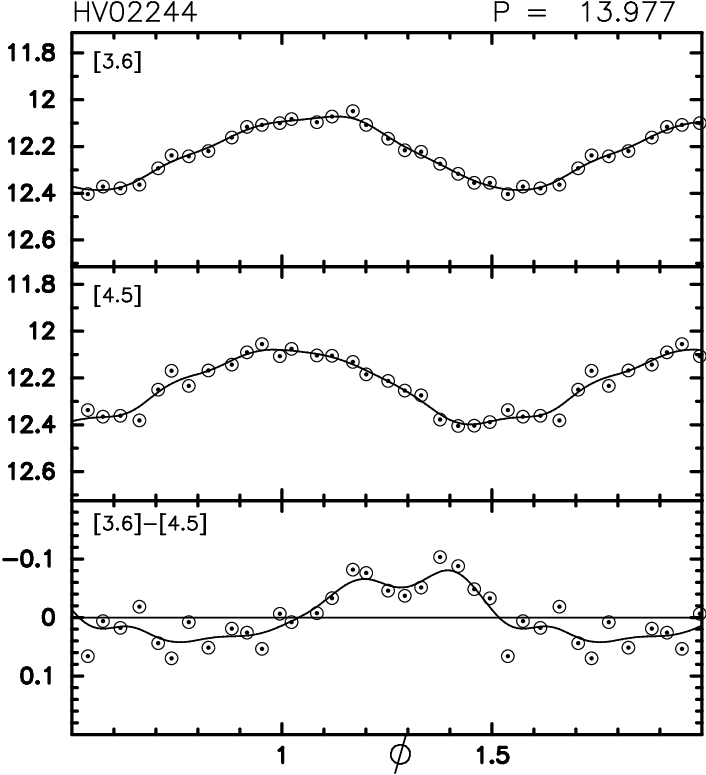} & 
\includegraphics[width=50mm,angle=270]{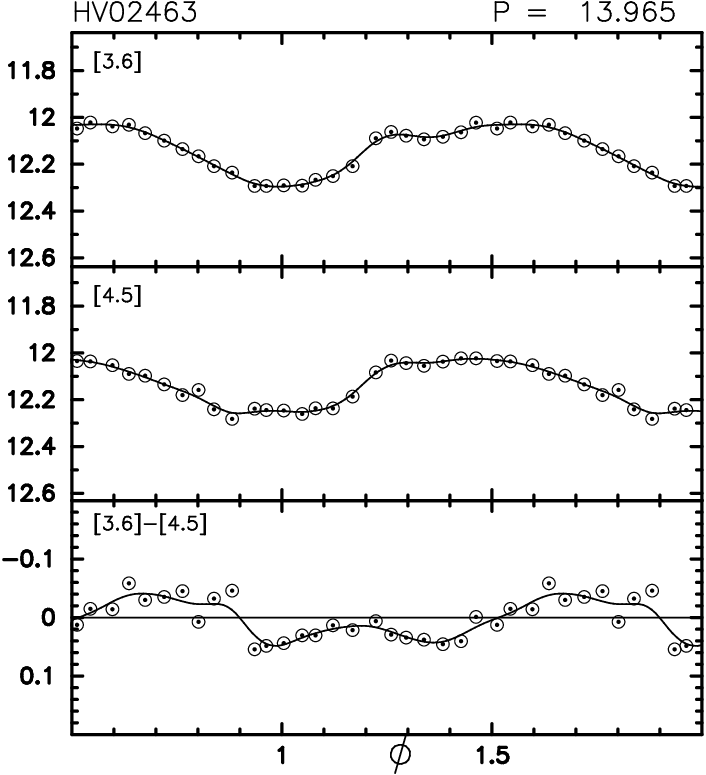} \\ 
\includegraphics[width=50mm,angle=270]{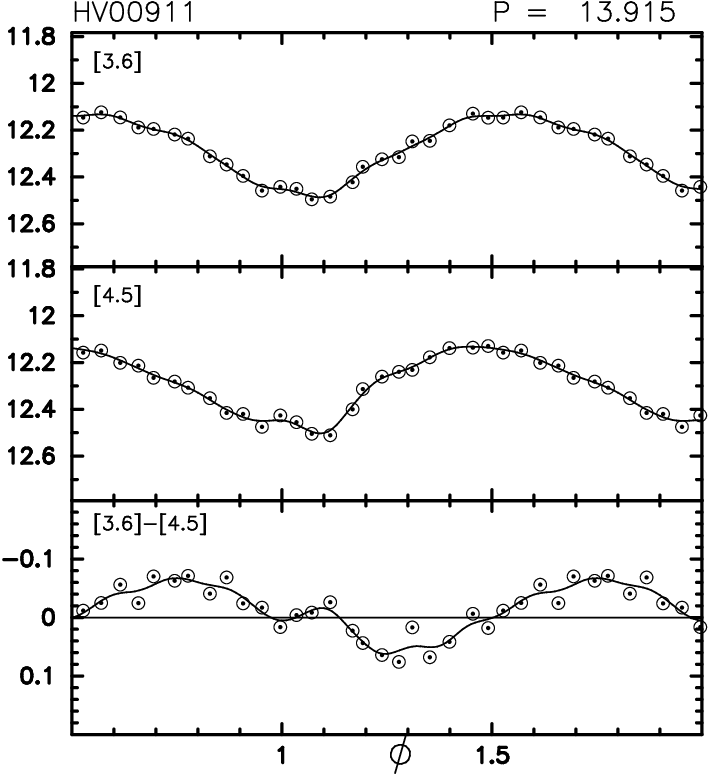} & 
 \includegraphics[width=50mm,angle=270]{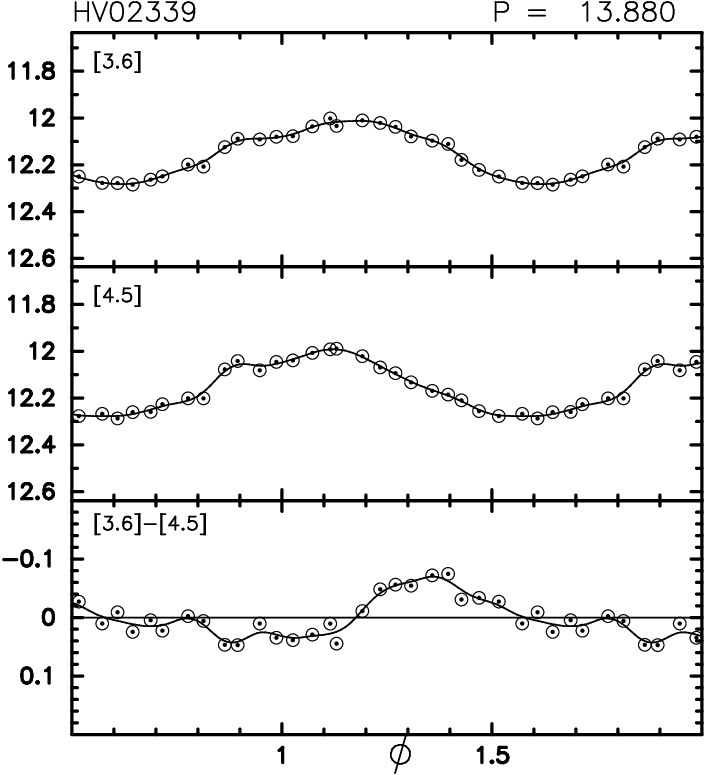} &
\includegraphics[width=50mm,angle=270]{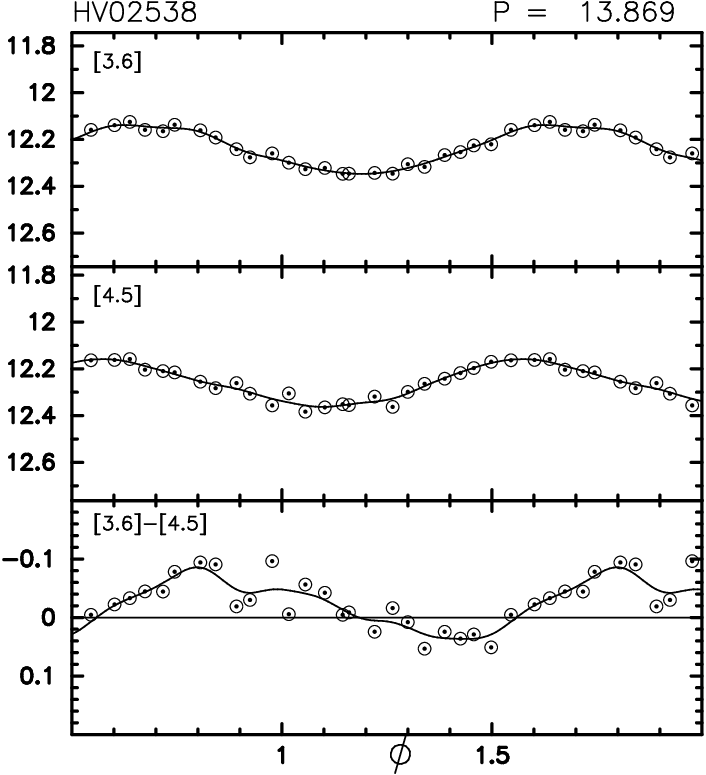} \\
\includegraphics[width=50mm,angle=270]{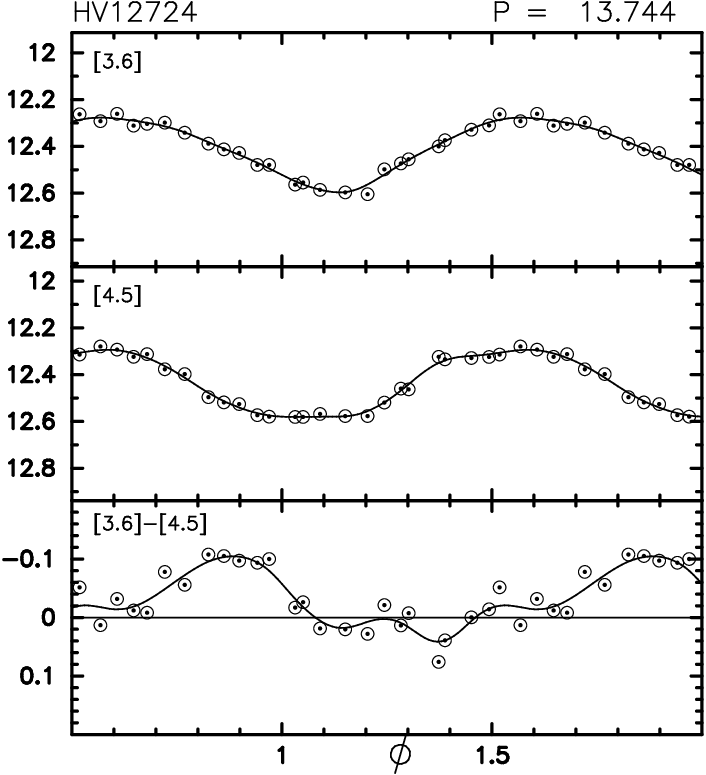} & 
\includegraphics[width=50mm,angle=270]{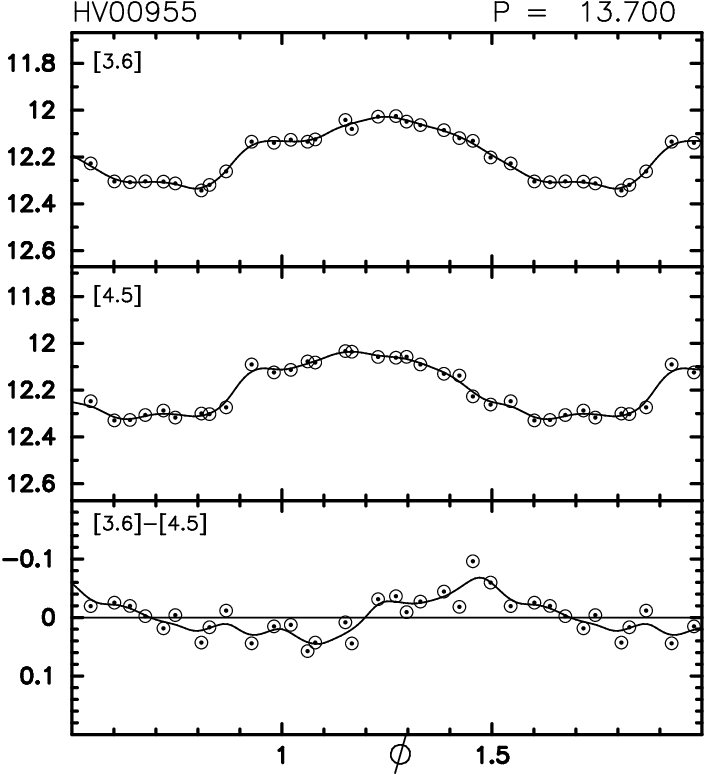} & 
\includegraphics[width=50mm,angle=270]{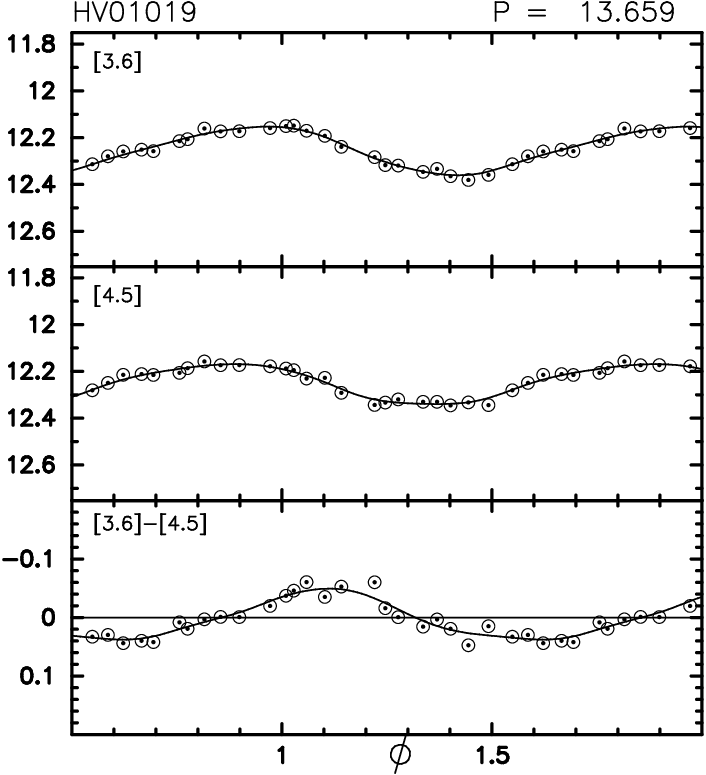} \\ 
\end{array}$ 
\end{center} 
\end{figure}
\begin{figure} 
 \begin{center}$ 
 \begin{array}{ccc} 
\includegraphics[width=50mm,angle=270]{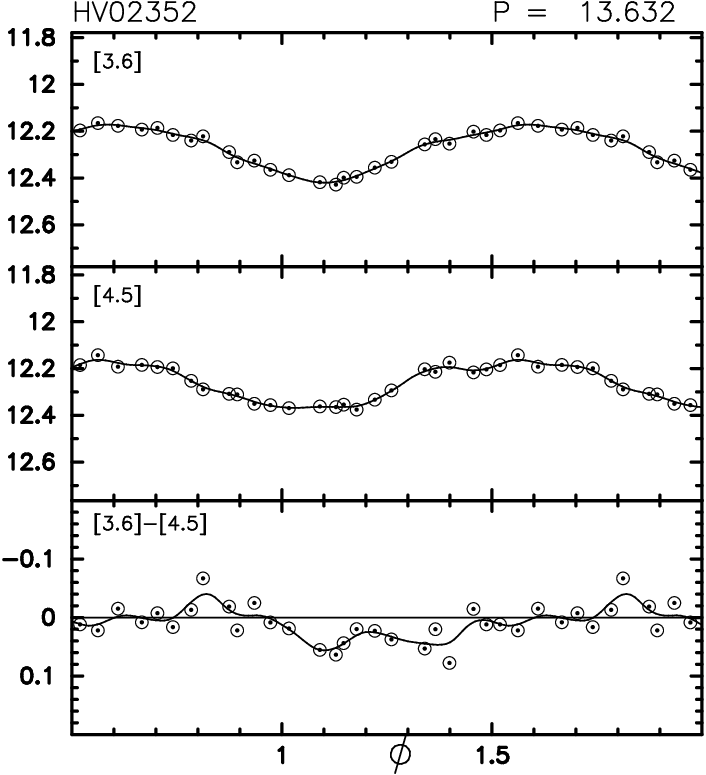} & 
\includegraphics[width=50mm,angle=270]{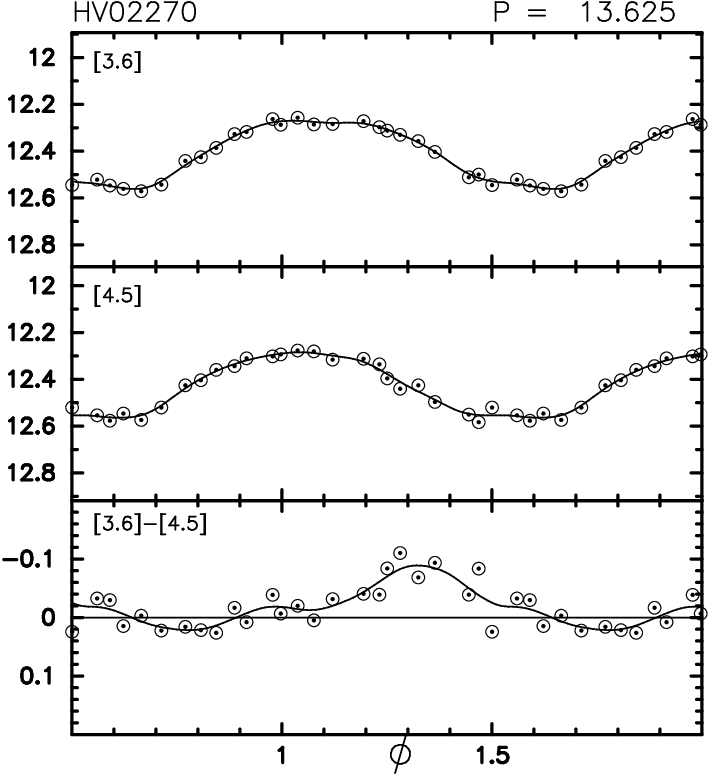} & 
\includegraphics[width=50mm,angle=270]{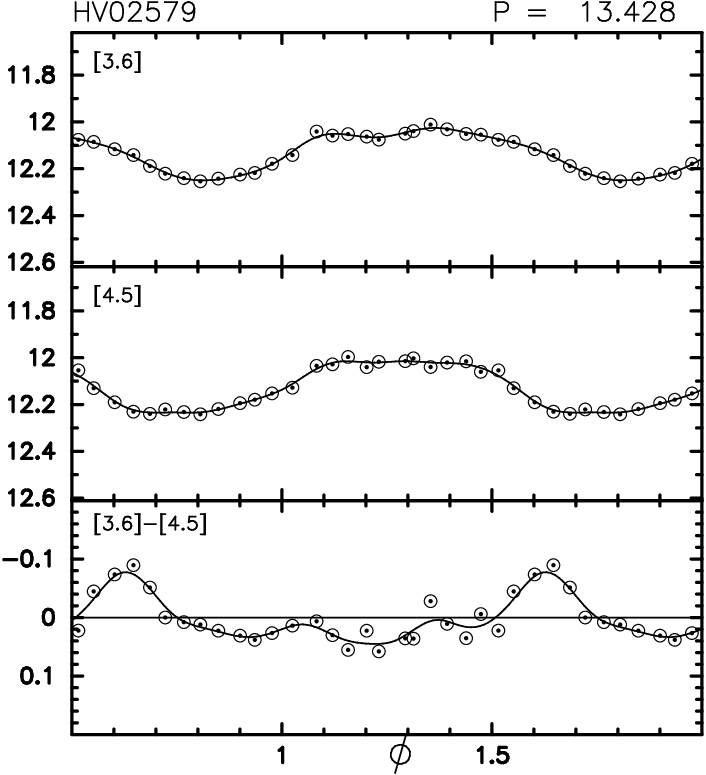} \\ 
\includegraphics[width=50mm,angle=270]{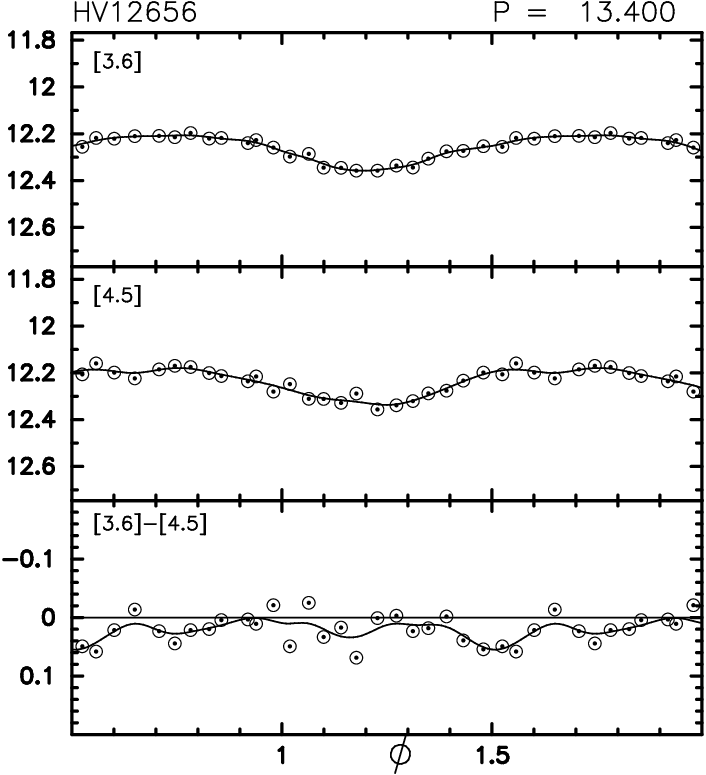} & 
\includegraphics[width=50mm,angle=270]{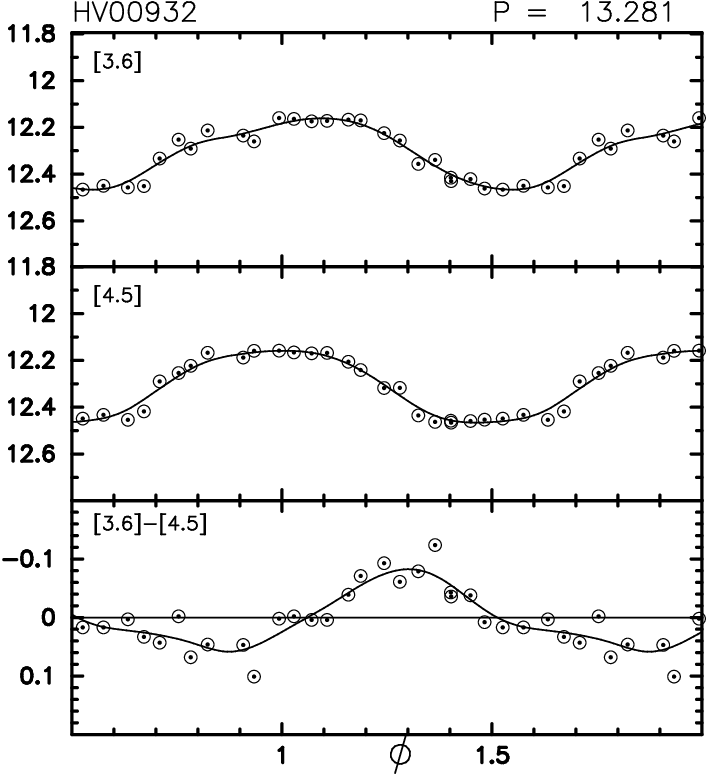} & 
\includegraphics[width=50mm,angle=270]{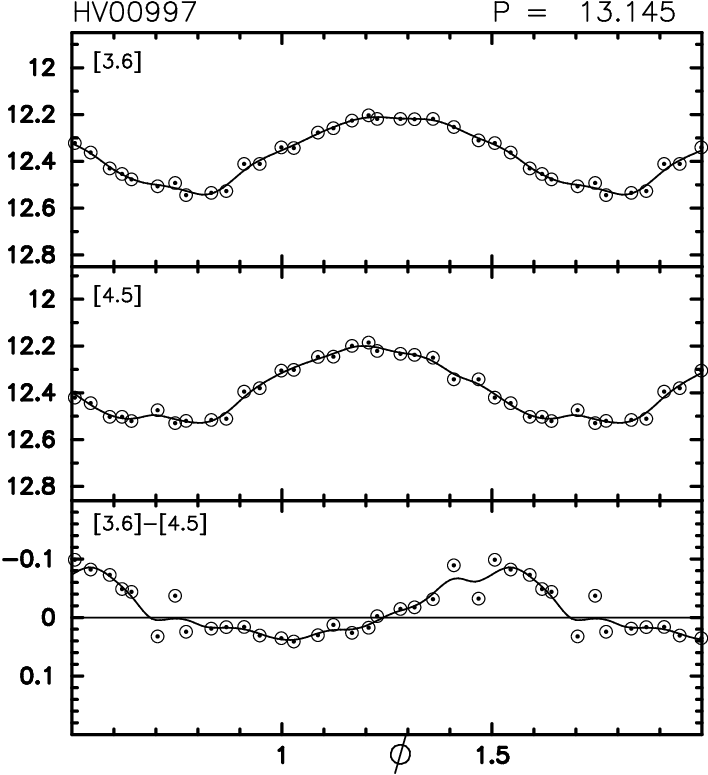} \\ 
\includegraphics[width=50mm,angle=270]{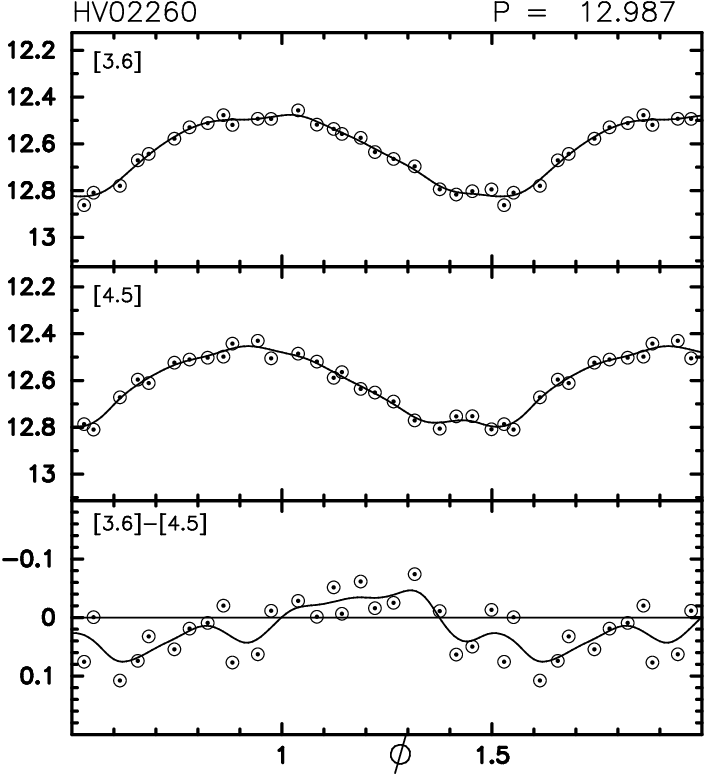} & 
\includegraphics[width=50mm,angle=270]{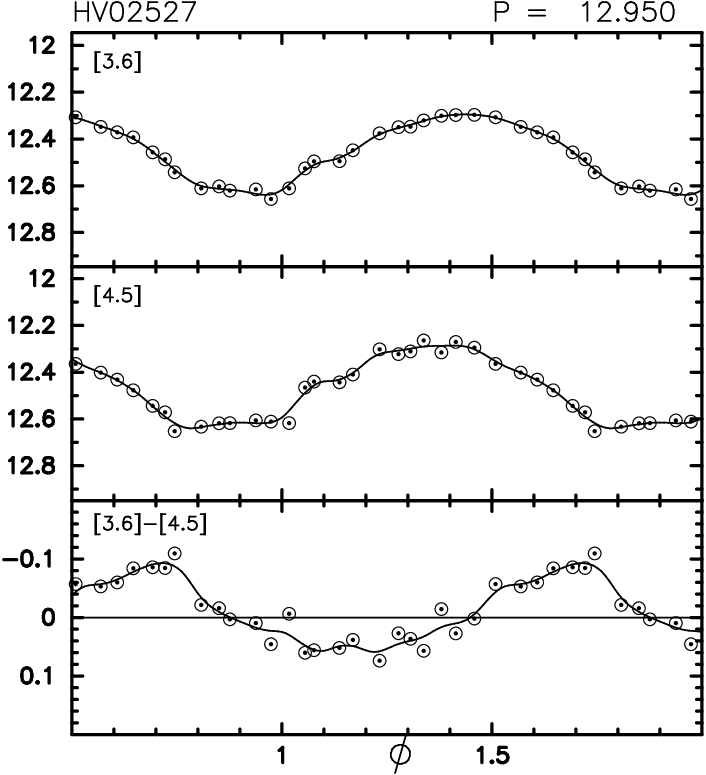} & 
\includegraphics[width=50mm,angle=270]{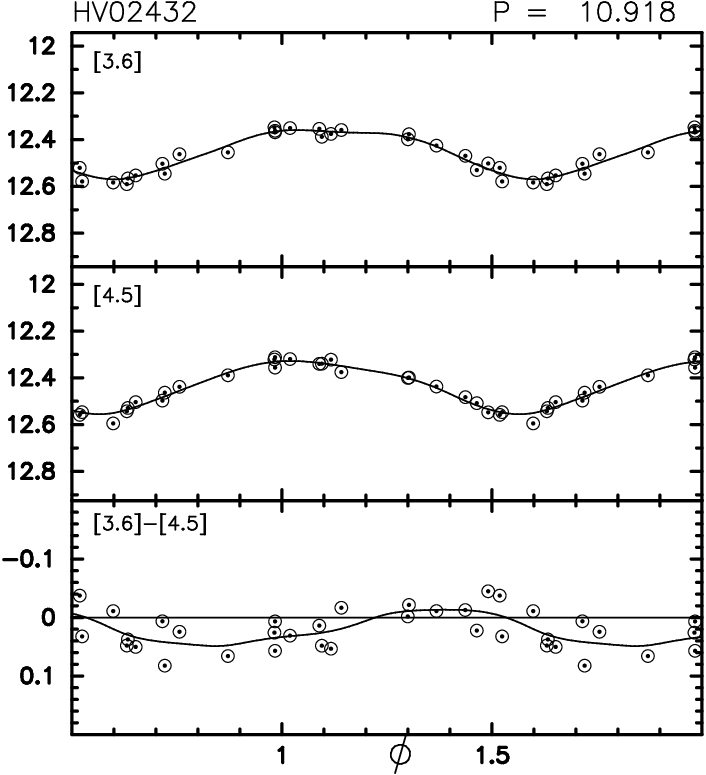} \\ 
\end{array}$ 
\end{center} 
\end{figure}
\begin{figure} 
 \begin{center}$ 
 \begin{array}{ccc} 
\includegraphics[width=50mm,angle=270]{f2_73.eps} & 
\includegraphics[width=50mm,angle=270]{f2_74.eps} & 
\includegraphics[width=50mm,angle=270]{f2_75.eps} \\ 
\includegraphics[width=50mm,angle=270]{f2_76.eps} & 
\includegraphics[width=50mm,angle=270]{f2_77.eps} & 
\includegraphics[width=50mm,angle=270]{f2_78.eps} \\ 
\includegraphics[width=50mm,angle=270]{f2_79.eps} & 
\includegraphics[width=50mm,angle=270]{f2_80.eps} & 
\includegraphics[width=50mm,angle=270]{f2_81.eps} \\ 
 \end{array}$ 
\end{center} 
\end{figure}
\begin{figure}
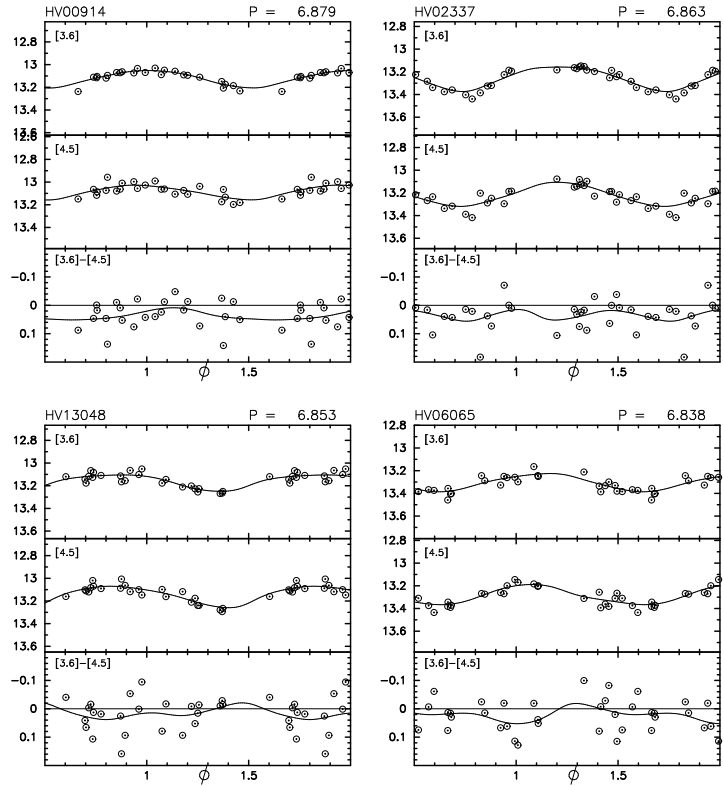
 
 \begin{center}$ 
 \begin{array}{cc} 
\includegraphics[width=50mm,angle=270]{f2_82.eps} & 
\includegraphics[width=50mm,angle=270]{f2_83.eps} \\ 
\includegraphics[width=50mm,angle=270]{f2_84.eps} & 
\includegraphics[width=50mm,angle=270]{f2_85.eps} \\ 
\end{array}$

\caption{IRAC light curves of the LMC Cepheid sample, in order of decreasing period. Point sizes are comparable to the uncertainties in the $[3.6]$ (top) and $[4.5]$ (middle) panels. The bottom panels show the variation of the IRAC color with phase. Upwards (more negative) in color corresponds to greater CO absorption as discussed in Section~\ref{sec:co_absorption}.}
\label{fig:light_curves}
 \end{center}
\end{figure}

\begin{figure}
 \begin{center}
  \includegraphics[width=150mm]{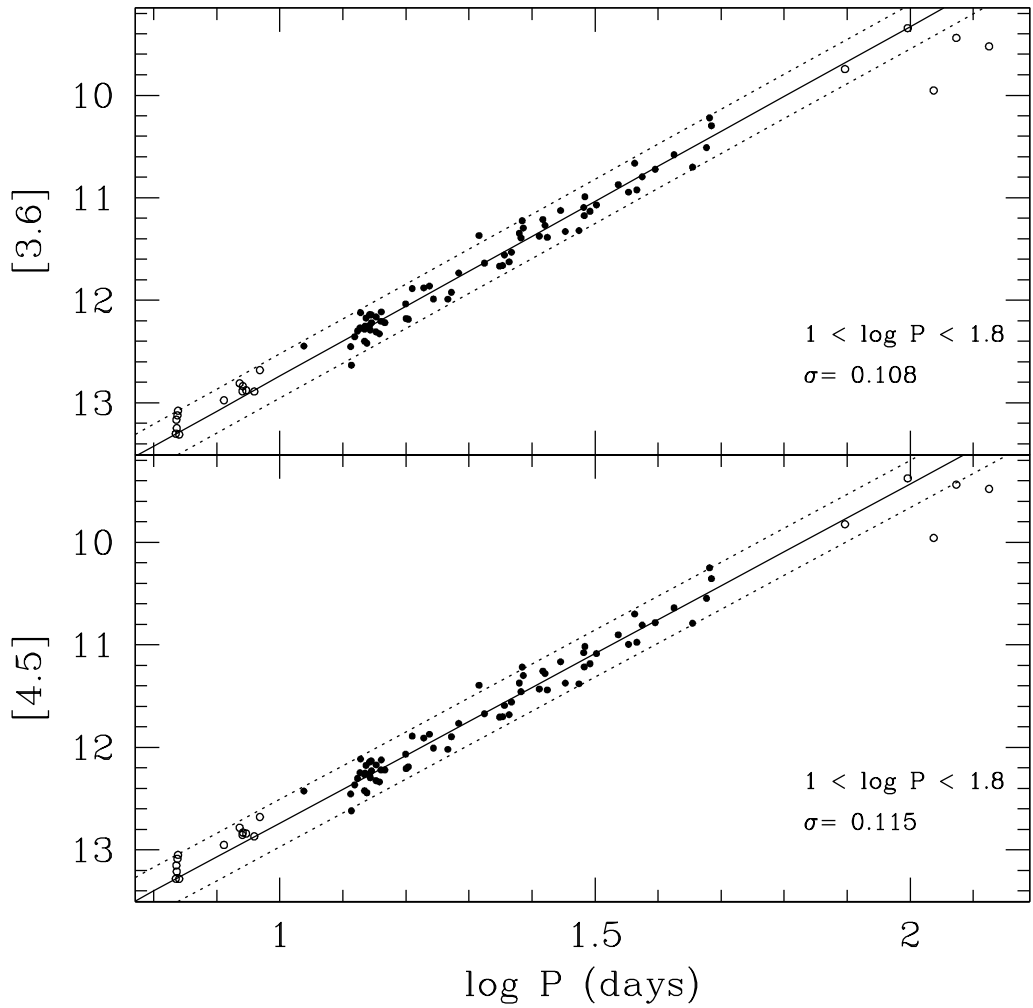}
\caption{LMC mid--infrared PL relations. The solid lines indicate the relations given in Table ~\ref{tab:pl_relations}, cut at 10 and 60 days, i.e. the open circles are not included in the fit. The dashed lines show $\pm 2 \sigma$. Error bars are consistent with the size of the points.}
\label{fig:pl_relations}
 \end{center}
\end{figure}

\begin{figure}
\begin{center}
\includegraphics[width=150mm]{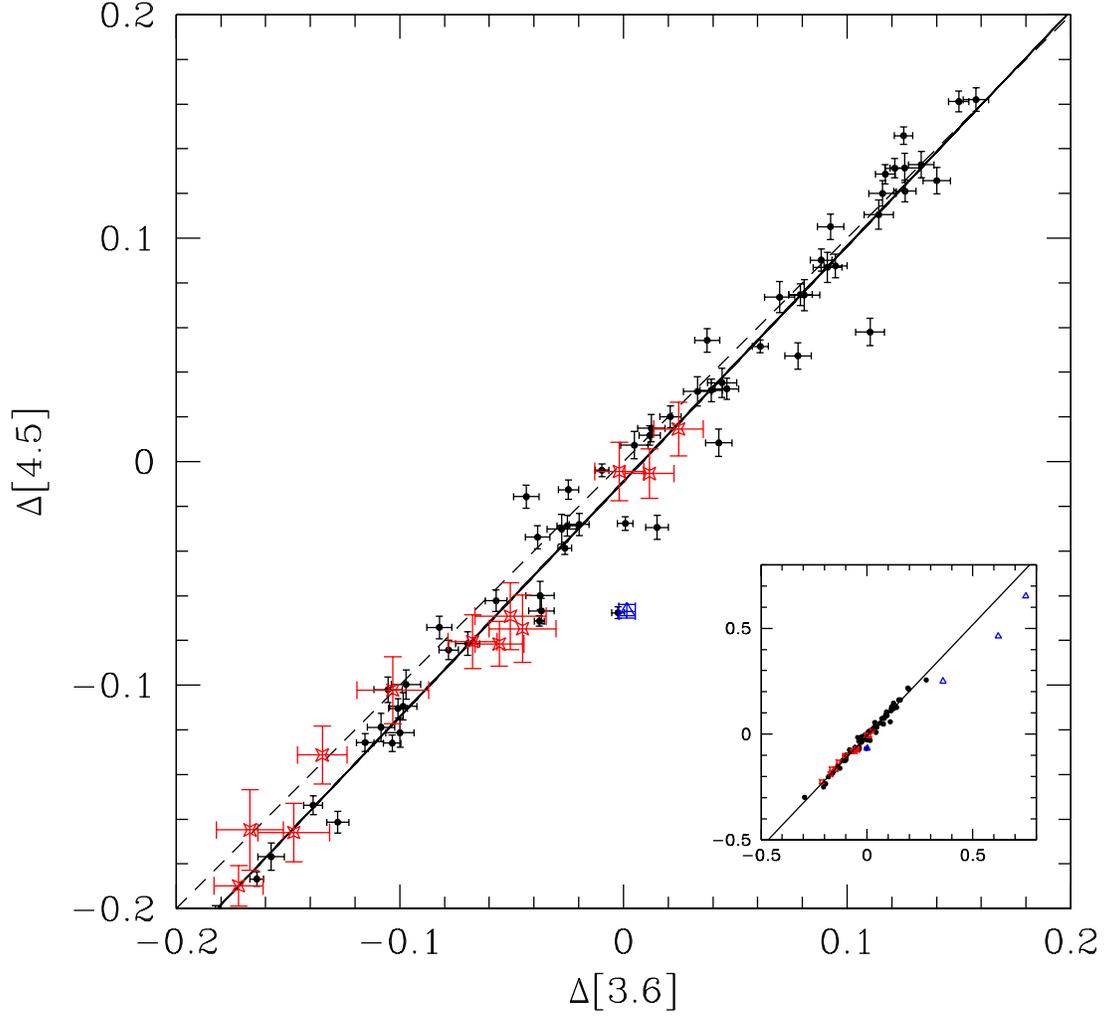}
\caption{Magnitude residuals from the [3.6] and [4.5] period--luminosity relations, with short period ($P < 10$ days) Cepheids as red open stars, long period ($P > 100$ days) as blue open triangles, and the PL fitting sample as black filled circles. The inset plot shows the whole sample; the main plot is cut at $\pm0.2$ mag to show an expanded view of the details. The dashed line indicates $\Delta[3.6] = \Delta[4.5]$, the solid line is a fit to the black points. The deviation from a one-to-one fit is believed to be due to CO pushing the [4.5] points further from the adopted PL. }
\label{fig:pl_residuals}
\end{center}
\end{figure}

\begin{figure}
\begin{center}
\includegraphics[width=150mm]{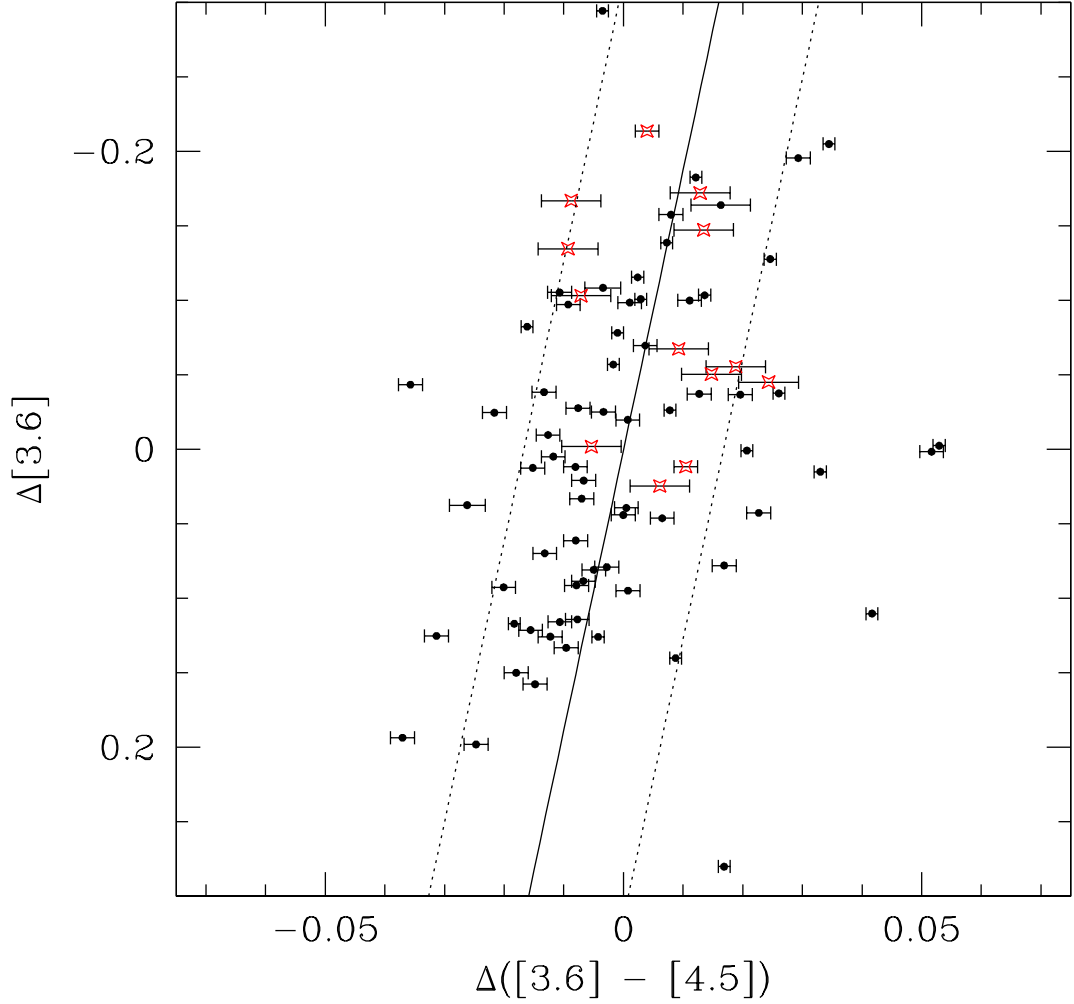}
\caption{Color and magnitude residuals from the period luminosity relation. The dashed lines show the $1\sigma$ dispersion. Excluding the longest period ($\log P > 2$) stars, the color dispersion around the [3.6] ridge line is 0.018 mag. Assuming that the instability strip has a regularly filled, rectangular distribution, the width will be $\sqrt{12} \sigma$. This gives the color width of the instability strip as 0.062 mag. }
\label{fig:colour_mag_residuals}
\end{center}
\end{figure}

\begin{figure}
 \begin{center}
  \includegraphics[width=150mm,angle=270]{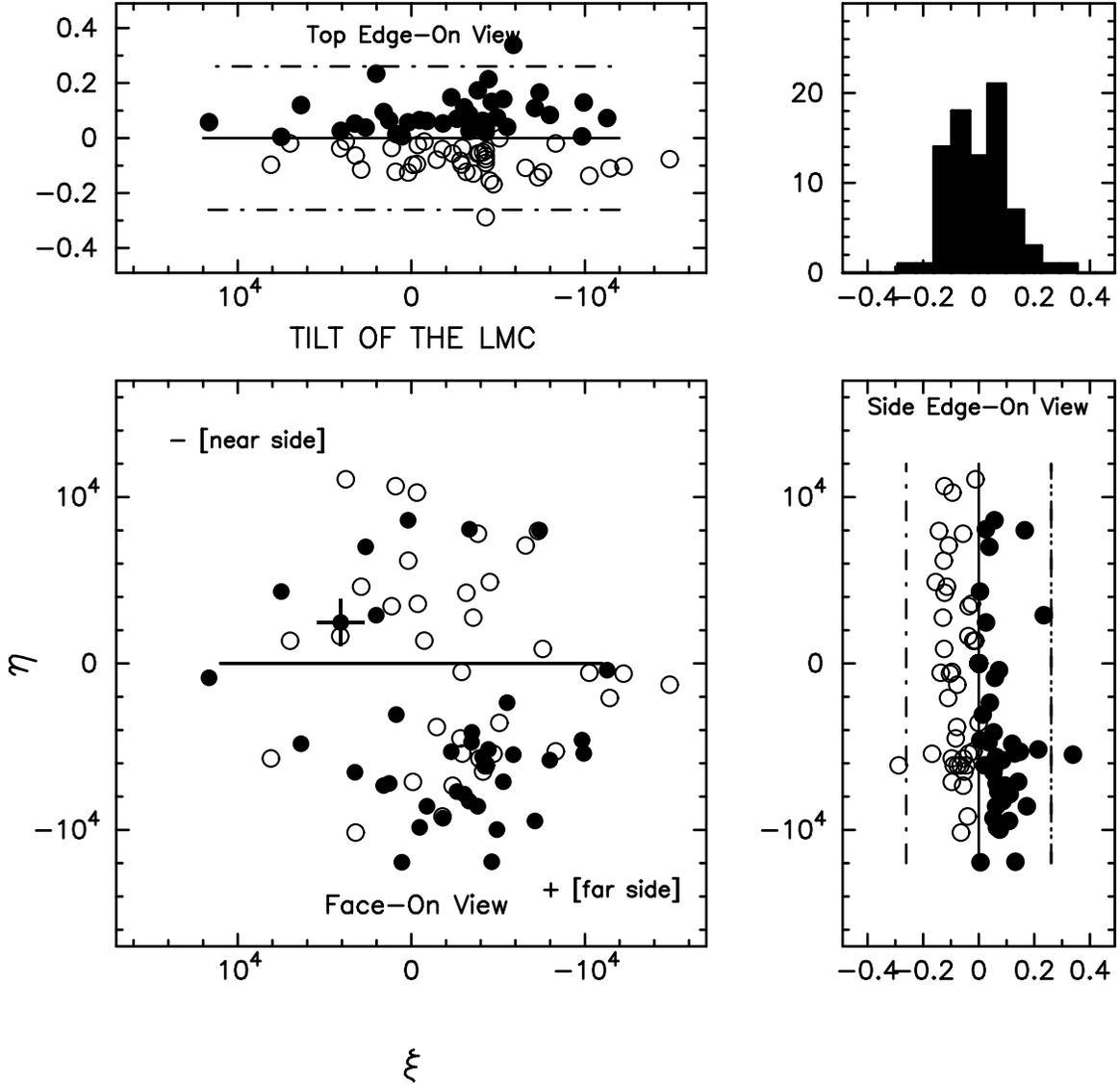}
\caption{Positions of the Cepheids within the LMC. Open circles show Cepheids which lie above the PL relation (i.e. are brighter than average), filled circles show Cepheids which lie below it. The bottom left plot shows their positions relative to the center of the LMC, with the coordinate system rotated such that the line of nodes is horizontal. The top left and bottom right plots show the deviations from the PL relation separated into the $\xi$ and $\eta$ components. The top right plot shows the histogram of the size of the PL deviations. The dispersion in the PL is minimized by assuming a tilt of $28^{\circ}$ in $\eta$ relative to the plane of the sky. }
\label{fig:tilt_plot}
 \end{center}
\end{figure}

\begin{figure}
\begin{center}
\includegraphics[width=150mm]{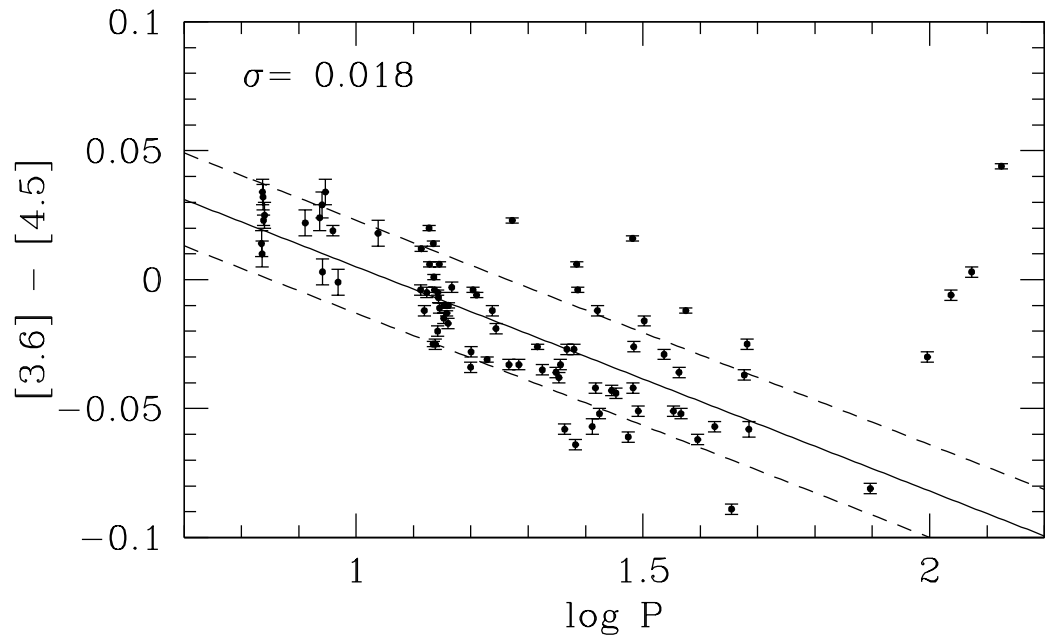}
\caption{Period--color relation for our LMC sample. The solid line shows the PC relation of Equation~\ref{eqn:pc_relation}, with the dashed lines showing $\pm 1 \sigma$. The trend apparently reverses for the very longest period ($\log P \geq 2$) objects.}
\label{fig:pc_relation}
\end{center}
\end{figure}

\begin{figure}
\begin{center}
\includegraphics[width=150mm]{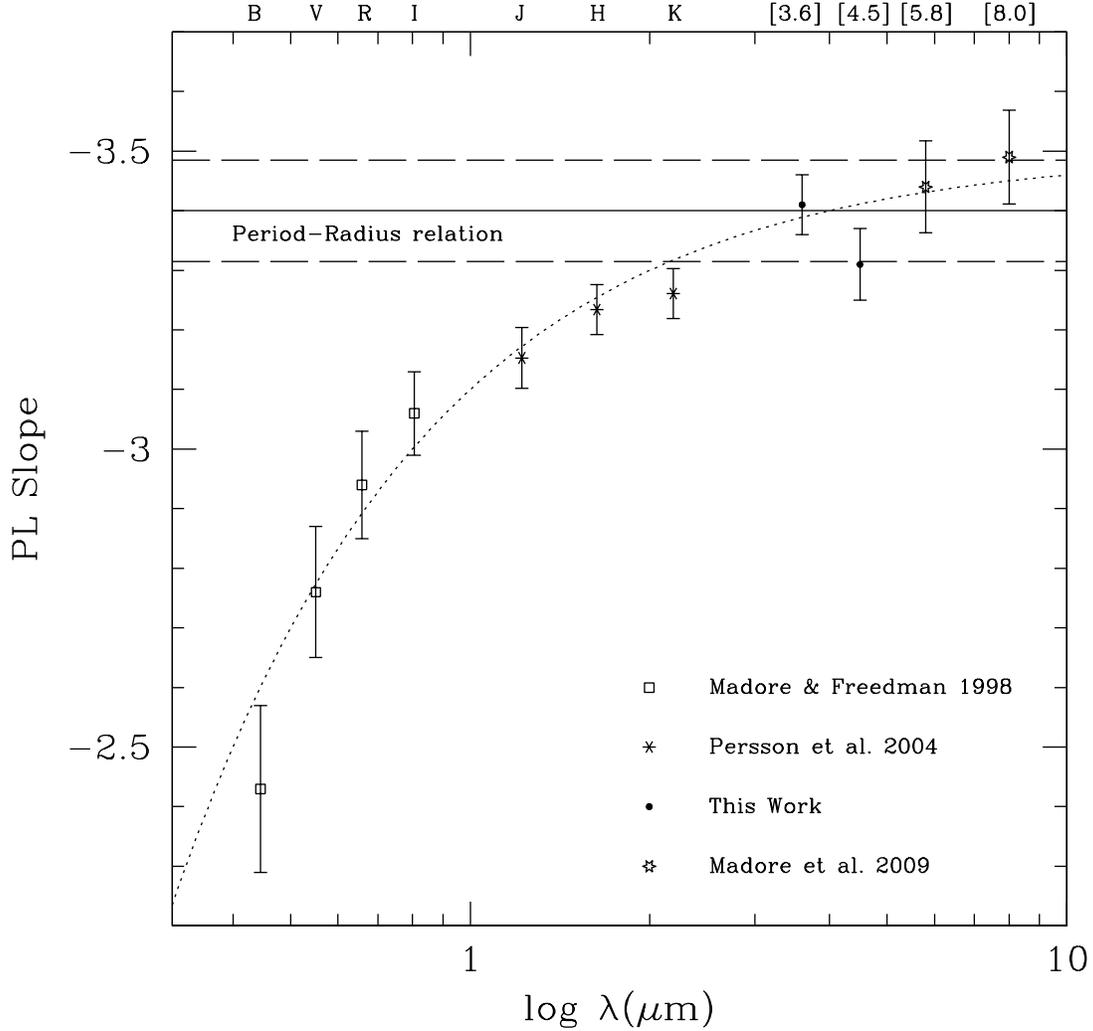}
\caption{Relationship between effective wavelength and the slope of the Cepheid PL relation. The PL slopes are taken from \citet{1998salg.conf..263M} ($BVRI$), P04 ($JHK$), this work ([3.6], [4.5]) and \citet{2009ApJ...695..988M} ([5.8], [8.0]). There is an asymptotic trend towards $-3.45$, which is consistent with the theoretical prediction of a slope of $-3.4$ for the period--radius relation. However, the $[4.5]$ (and to a smaller extent, $K$) slope is clearly discrepant. We attribute this to the CO bandhead that suppresses flux in the 4.5~$\mu$m. The [3.6] slope is not affected --- it follows the trend of the other wavelengths --- and is therefore appropriate to use for distance determinations. }
\label{fig:pl_slopes}
\end{center}
\end{figure}

\clearpage

\begin{deluxetable}{l c c c c c c c}
\tablecolumns{6}
\tablewidth{0pc}
\tablecaption{Magnitudes for LMC Cepheids. \label{tab:mag_results}
}
\tablehead{
\colhead{Cepheid} & \colhead{Period} & \colhead{[3.6]} &
\colhead{$\sigma_{[3.6]}$\tablenotemark{a}} & \colhead{[4.5]} &
\colhead{$\sigma_{[4.5]}$\tablenotemark{a}}  & \colhead{$[3.6]-[4.5]$} & \colhead{$\sigma_{[3.6]-[4.5]}$\tablenotemark{a}}\\
 & \colhead{(days)} & \colhead {(mag)} & & \colhead {(mag)} & &
 \colhead{(mag)} & }
\startdata
HV00872 & 29.820 & 11.320 & 0.005 & 11.381 & 0.005 & -0.061 & 0.002 \\
HV00873 & 34.426 & 10.872 & 0.006 & 10.903 & 0.006 & -0.029 & 0.002 \\
HV00875 & 30.338 & 11.094 & 0.003 & 11.077 & 0.003 & 0.016 & 0.001 \\
HV00876 & 22.716 & 11.558 & 0.006 & 11.592 & 0.007 & -0.033 & 0.002 \\
HV00877 & 45.155 & 10.701 & 0.004 & 10.790 & 0.004 & -0.089 & 0.002 \\
HV00878 & 23.306 & 11.531 & 0.007 & 11.559 & 0.007 & -0.027 & 0.002 \\
HV00879 & 36.840 & 10.923 & 0.007 & 10.976 & 0.007 & -0.052 & 0.002 \\
HV00881 & 35.740 & 10.945 & 0.007 & 10.996 & 0.007 & -0.051 & 0.002 \\
HV00882 & 31.786 & 11.070 & 0.006 & 11.086 & 0.006 & -0.016 & 0.002 \\
HV00883 & 133.400 & 9.523 & 0.011 & 9.479 & 0.010 & 0.044 & 0.001 \\
HV00885 & 20.703 & 11.368 & 0.003 & 11.395 & 0.003 & -0.026 & 0.001 \\
HV00886 & 23.983 & 11.346 & 0.006 & 11.373 & 0.006 & -0.027 & 0.002 \\
HV00887 & 14.486 & 12.113 & 0.004 & 12.123 & 0.004 & -0.010 & 0.001 \\
HV00889 & 25.797 & 11.374 & 0.006 & 11.432 & 0.005 & -0.057 & 0.003 \\
HV00891 & 17.272 & 11.861 & 0.005 & 11.873 & 0.005 & -0.012 & 0.002 \\
HV00892 & 15.990 & 12.185 & 0.006 & 12.191 & 0.006 & -0.004 & 0.001 \\
HV00893 & 21.115 & 11.638 & 0.006 & 11.673 & 0.006 & -0.035 & 0.002 \\
HV00899 & 31.047 & 11.132 & 0.007 & 11.185 & 0.007 & -0.051 & 0.002 \\
HV00900 & 47.515 & 10.510 & 0.006 & 10.547 & 0.006 & -0.037 & 0.002 \\
HV00901 & 18.470 & 11.989 & 0.006 & 12.020 & 0.005 & -0.033 & 0.002 \\
HV00902 & 26.340 & 11.269 & 0.006 & 11.281 & 0.006 & -0.012 & 0.002 \\
HV00904 & 30.405 & 11.174 & 0.007 & 11.216 & 0.007 & -0.042 & 0.002 \\
HV00909 & 37.566 & 10.795 & 0.005 & 10.808 & 0.005 & -0.012 & 0.001 \\
HV00911 & 13.915 & 12.290 & 0.005 & 12.297 & 0.005 & -0.007 & 0.002 \\
HV00914 & 6.879 & 13.122 & 0.011 & 13.088 & 0.009 & 0.032 & 0.005 \\
HV00932 & 13.281 & 12.300 & 0.004 & 12.304 & 0.005 & -0.005 & 0.002 \\
HV00953 & 48.048 & 10.220 & 0.005 & 10.249 & 0.005 & -0.025 & 0.002 \\
HV00955 & 13.700 & 12.173 & 0.004 & 12.177 & 0.004 & -0.004 & 0.001 \\
HV00971 & 9.297 & 12.681 & 0.015 & 12.680 & 0.018 & -0.001 & 0.005 \\
HV00997 & 13.145 & 12.356 & 0.005 & 12.367 & 0.005 & -0.012 & 0.002 \\
HV01002 & 30.472 & 10.990 & 0.006 & 11.017 & 0.007 & -0.026 & 0.002 \\
HV01003 & 24.339 & 11.295 & 0.005 & 11.300 & 0.005 & -0.004 & 0.001 \\
HV01005 & 18.716 & 11.922 & 0.007 & 11.897 & 0.006 & 0.023 & 0.001 \\
HV01006 & 14.217 & 12.162 & 0.005 & 12.172 & 0.005 & -0.010 & 0.001 \\
HV01013 & 24.131 & 11.392 & 0.006 & 11.458 & 0.005 & -0.064 & 0.002 \\
HV01019 & 13.659 & 12.252 & 0.003 & 12.253 & 0.003 & 0.001 & 0.001 \\
HV01023 & 26.559 & 11.386 & 0.006 & 11.441 & 0.006 & -0.052 & 0.002 \\
HV02244 & 13.977 & 12.219 & 0.004 & 12.230 & 0.005 & -0.011 & 0.002 \\
HV02251 & 27.887 & 11.124 & 0.007 & 11.166 & 0.007 & -0.043 & 0.002 \\
HV02257 & 39.395 & 10.722 & 0.006 & 10.784 & 0.006 & -0.062 & 0.002 \\
HV02260 & 12.987 & 12.633 & 0.005 & 12.620 & 0.005 & 0.012 & 0.001 \\
HV02262 & 15.832 & 12.035 & 0.004 & 12.067 & 0.004 & -0.034 & 0.002 \\
HV02270 & 13.625 & 12.399 & 0.004 & 12.424 & 0.004 & -0.025 & 0.001 \\
HV02279 & 6.895 & 13.077 & 0.013 & 13.050 & 0.014 & 0.023 & 0.002 \\
HV02282 & 14.677 & 12.218 & 0.005 & 12.221 & 0.005 & -0.003 & 0.002 \\
HV02291 & 22.317 & 11.667 & 0.006 & 11.706 & 0.006 & -0.036 & 0.002 \\
HV02294 & 36.552 & 10.663 & 0.006 & 10.700 & 0.006 & -0.036 & 0.002 \\
HV02324 & 14.466 & 12.205 & 0.005 & 12.221 & 0.004 & -0.017 & 0.002 \\
HV02337 & 6.863 & 13.247 & 0.016 & 13.212 & 0.015 & 0.034 & 0.005 \\
HV02338 & 42.200 & 10.580 & 0.007 & 10.640 & 0.007 & -0.057 & 0.002 \\
HV02339 & 13.880 & 12.139 & 0.004 & 12.143 & 0.004 & -0.005 & 0.001 \\
HV02352 & 13.632 & 12.282 & 0.003 & 12.267 & 0.003 & 0.014 & 0.001 \\
HV02369 & 48.377 & 10.297 & 0.006 & 10.355 & 0.006 & -0.058 & 0.003 \\
HV02405 & 6.924 & 13.309 & 0.011 & 13.283 & 0.012 & 0.025 & 0.005 \\
HV02432 & 10.918 & 12.446 & 0.015 & 12.427 & 0.016 & 0.018 & 0.005 \\
HV02447 & 118.350 & 9.440 & 0.006 & 9.438 & 0.005 & 0.003 & 0.002 \\
HV02463 & 13.965 & 12.142 & 0.004 & 12.134 & 0.004 & 0.006 & 0.001 \\
HV02527 & 12.950 & 12.452 & 0.005 & 12.456 & 0.005 & -0.004 & 0.002 \\
HV02538 & 13.869 & 12.246 & 0.003 & 12.266 & 0.003 & -0.020 & 0.002 \\
HV02549 & 16.220 & 11.885 & 0.004 & 11.891 & 0.004 & -0.006 & 0.001 \\
HV02579 & 13.428 & 12.121 & 0.003 & 12.114 & 0.004 & 0.006 & 0.001 \\
HV02580 & 16.918 & 11.879 & 0.006 & 11.910 & 0.005 & -0.031 & 0.001 \\
HV02733 & 8.722 & 12.887 & 0.011 & 12.855 & 0.010 & 0.029 & 0.005 \\
HV02749 & 23.105 & 11.625 & 0.004 & 11.682 & 0.004 & -0.058 & 0.002 \\
HV02793 & 19.220 & 11.734 & 0.006 & 11.767 & 0.005 & -0.033 & 0.002 \\
HV02827 & 78.802 & 9.744 & 0.004 & 9.824 & 0.003 & -0.081 & 0.002 \\
HV02836 & 17.528 & 11.988 & 0.005 & 12.008 & 0.005 & -0.019 & 0.002 \\
HV02854 & 8.635 & 12.810 & 0.016 & 12.785 & 0.013 & 0.024 & 0.005 \\
HV02883 & 108.970 & 9.953 & 0.010 & 9.958 & 0.009 & -0.006 & 0.002 \\
HV05497 & 99.100 & 9.345 & 0.004 & 9.376 & 0.003 & -0.030 & 0.002 \\
HV05655 & 14.211 & 12.308 & 0.005 & 12.325 & 0.005 & -0.015 & 0.002 \\
HV06065 & 6.838 & 13.301 & 0.011 & 13.282 & 0.013 & 0.014 & 0.005 \\
HV06098 & 24.237 & 11.224 & 0.003 & 11.218 & 0.003 & 0.006 & 0.001 \\
HV08036 & 28.369 & 11.329 & 0.006 & 11.374 & 0.006 & -0.044 & 0.002 \\
HV12452 & 8.736 & 12.837 & 0.016 & 12.832 & 0.015 & 0.003 & 0.005 \\
HV12471 & 15.863 & 12.178 & 0.004 & 12.208 & 0.004 & -0.028 & 0.002 \\
HV12505 & 14.389 & 12.327 & 0.005 & 12.338 & 0.005 & -0.013 & 0.001 \\
HV12656 & 13.400 & 12.269 & 0.002 & 12.248 & 0.002 & 0.020 & 0.001 \\
HV12700 & 8.153 & 12.975 & 0.011 & 12.953 & 0.012 & 0.022 & 0.005 \\
HV12717 & 8.843 & 12.877 & 0.015 & 12.842 & 0.015 & 0.034 & 0.005 \\
HV12724 & 13.744 & 12.419 & 0.004 & 12.444 & 0.005 & -0.025 & 0.002 \\
HV12815 & 26.135 & 11.212 & 0.006 & 11.257 & 0.006 & -0.042 & 0.002 \\
HV12816 & 9.108 & 12.890 & 0.011 & 12.869 & 0.011 & 0.019 & 0.002 \\
HV13048 & 6.853 & 13.165 & 0.011 & 13.152 & 0.013 & 0.010 & 0.005 \\
P71-U01 & 22.558 & 11.661 & 0.007 & 11.702 & 0.007 & -0.038 & 0.002 \\
\enddata
\tablenotetext{a}{For Cepheids with P $\leq 12$ days the uncertainties scale with 1/$\sqrt{N}$ rather than $1/N$ as the observations of short period Cepheids were not phase locked.}
\end{deluxetable}

\begin{deluxetable}{l c c c c}
\tablecolumns{6}
\tablewidth{0pc}
\tablecaption{Photometry of HV00872. \label{tab:example_photometry_hv872}
}
\tablehead{
\colhead{MJD\tablenotemark{b}} & \colhead{[3.6]\tablenotemark{b}} & \colhead{$\sigma_{[3.6]}$} & \colhead{[4.5]\tablenotemark{b}} &
\colhead{$\sigma_{[4.5]}$}  \\
  & \colhead {(mag)} & & \colhead {(mag)} & }
\startdata
55133.665 & 11.538 & 0.013 & 11.633 & 0.015 \\
55134.756 & 11.568 & 0.013 & 11.605 & 0.014 \\
55136.253 & 11.510 & 0.013 & 11.498 & 0.013 \\
55136.953 & 11.485 & 0.012 & 11.471 & 0.014 \\
55138.621 & 11.396 & 0.013 & 11.372 & 0.013 \\
55139.561 & 11.378 & 0.012 & 11.330 & 0.013 \\
55140.963 & 11.334 & 0.011 & 11.296 & 0.013 \\
55141.986 & 11.268 & 0.011 & 11.276 & 0.012 \\
55143.049 & 11.244 & 0.012 & 11.255 & 0.012 \\
55144.443 & 11.205 & 0.011 & 11.235 & 0.013 \\
55146.159 & 11.182 & 0.011 & 11.232 & 0.012 \\
55146.916 & 11.162 & 0.011 & 11.235 & 0.012 \\
55148.340 & 11.165 & 0.011 & 11.243 & 0.012 \\
55149.495 & 11.167 & 0.011 & 11.267 & 0.014 \\
55151.036 & 11.171 & 0.011 & 11.290 & 0.012 \\
55151.989 & 11.187 & 0.011 & 11.292 & 0.012 \\
55153.002 & 11.220 & 0.011 & 11.315 & 0.013 \\
55155.321 & 11.266 & 0.011 & 11.394 & 0.013 \\
55155.906 & 11.282 & 0.011 & 11.378 & 0.013 \\
55156.963 & 11.302 & 0.011 & 11.419 & 0.013 \\
55158.231 & 11.357 & 0.012 & 11.460 & 0.013 \\
55159.318 & 11.389 & 0.012 & 11.528 & 0.014 \\
55160.793 & 11.439 & 0.012 & 11.523 & 0.014 \\
55161.870 & 11.472 & 0.012 & 11.584 & 0.014 \\ \hline 
 \enddata
 \tablecomments{The Cepheid photometry is available in the electronic edition of the {\it Astrophysical Journal}.}
\tablenotetext{a}{MJD = JD $-$ 2,400,000.5 d}
\tablenotetext{b}{Each magnitude point corresponds to a flux weighted average of the five dither positions.}
\end{deluxetable}

\begin{table}
\begin{center}
\caption{Period--luminosity relations }
\label{tab:pl_relations}
\begin{tabular}{ l l c c c c c c} \\ \hline \hline
Band & Sample & $N$ &$a$ & $\sigma_{a}$ & $b$ & $\sigma_{b}$& $\sigma$\\ \hline
[3.6] & $10$d $ \leq P \leq 60$d & 67 & $-3.41$ & $0.08$ & $12.74$ & $0.03$ & $0.108$ \\ 
 & SAGE\tablenotemark{a} & 70 & $-3.40$ & $0.07$ & $12.73$ & $0.02$ & $0.135$ \\ 
 & No cut & 85 & $-3.16$ & $0.05$ & $12.67$ & $0.02$ & $0.134$ \\
 & 6 d $\leq P \leq 60$ d & 82 & $-3.31$ & $0.05$ & $12.74$ & $0.01$ & $0.105$ \\
 & $P \geq 10$ d & 72 & $-3.14$ & $0.07$ & $12.67$ & $0.03$ & $0.142$ \\ \hline
[4.5] & $10$d $ \leq P \leq 60$d & 67 & $-3.31$ & $0.08$ & $12.74$ & $0.03$ & $0.115$ \\ 
 & SAGE\tablenotemark{a}  & 70 & $-3.35$ & $0.07$ & $12.77$ & $0.03$ & $0.141$ \\
& No cut & 85 & $-3.10$ & $0.05$ & $12.67$ & $0.02$ & $0.131$ \\ 	
 & 6 d $\leq P \leq 60$ d & 82 & $-3.22$ & $0.05$ & $12.70$ & $0.02$ & $0.112$ \\
 & $P \geq 10$ d & 72 & $-3.11$ & $0.07$ & $12.68$  & $0.03$ & $0.139$ \\ \hline
\end{tabular}
\tablenotetext{a}{The SAGE data from \citet{2009ApJ...695..988M} were re-fit to be consistent with the CHP analysis.}
\end{center}
\end{table}

\end{document}